\documentclass[12pt]{article}
\usepackage{amsmath}
\usepackage{amssymb}
\usepackage{graphicx}
\usepackage{verbatim}
\usepackage{color}
\usepackage{subfigure}
\usepackage{hyperref}
\usepackage{cite}
\usepackage{bm}
\usepackage{natbib}
\usepackage{soul}
\usepackage{array}
\newcolumntype{L}[1]{>{\raggedright\let\newline\\\arraybackslash\hspace{0pt}}m{#1}}
\newcolumntype{C}[1]{>{\centering\let\newline\\\arraybackslash\hspace{0pt}}m{#1}}
\newcolumntype{R}[1]{>{\raggedleft\let\newline\\\arraybackslash\hspace{0pt}}m{#1}}

\title{Active-set algorithms based statistical inference for shape-restricted generalized additive Cox regression models}
\author{ Geng Deng, PhD\footnote{Corporate Risk, Wellsfargo \& Company, email:gengdeng@gmail.com} \and Guangning Xu, PhD\footnote{Corporate Risk, Wellsfargo \& Company, email:guangning.xu@wellsfargo.com} \and Xindong Wang, PhD\footnote{Corporate Risk, Wellsfargo \& Company, email: xindong.wang@wellsfargo.com} \and Qiang Fu, PhD\footnote{Corporate Risk, Wellsfargo \& Company, email:ken.fu@wellsfargo.com} \and Jing Qin, PhD\footnote{National Institute of Health: email:jingqin@niaid.nih.gov} }

\begin{document}
\maketitle

\begin{abstract}

Recently the shape-restricted inference has gained popularity in statistical and econometric literature in order to
relax the linear or quadratic covariate effect in regression analyses. The typical shape-restricted covariate effect includes monotonic increasing, decreasing, convexity or concavity. In this paper, we introduce the shape-restricted inference to the celebrated Cox regression model (SR-Cox), in which the covariate response is modeled as shape-restricted additive functions. The SR-Cox regression approximates the shape-restricted functions using a spline basis expansion with data driven choice of knots.
The underlying minimization of negative log-likelihood function is formulated as a convex optimization problem, which is solved with an active-set optimization algorithm. The highlight of this algorithm is that it eliminates the superfluous knots automatically.
When covariate effects include combinations of convex or concave terms with unknown forms and linear terms,
the most interesting finding is that SR-Cox produces accurate linear covariate effect estimates
which are comparable to the maximum partial likelihood estimates if indeed the forms are known.
 We conclude that concave or convex SR-Cox models could significantly improve nonlinear covariate response recovery and model goodness of fit.

\end{abstract}

\section{Introduction} \label{sec:intro}

As a natural extension of the parametric likelihood based inference, the shape-restricted inference has recently gained
attention in statistical and econometric literature.
Either from the perspective of the physical theory, or the biologic principle, or the econometric law, the shape-restricted regression occurs naturally.
For example, in data development analysis (DEA), \citet{banker:max} formulated the relation between DEA models and the estimation of monotone increasing concave production frontiers.  Based on the economic theory, \cite{matzkin:semi} proposed the inference for the utility functions
by imposing increasing and concave constraints.  \cite{luss.rosset.shahar:efficient} applied monotone regression to identify gene-gene interactions.
Basically, the shape-constrained inference is
classified in two categories: 1) the spline-based approach, in which the tuning parameters, such as the basis functions and the knots should be pre-specified and which may not be easy in practical applications, especially when the sample size is small or medium;
2) the nonparametric likelihood-based approach, in which the jumps of the baseline function are located in the observed data points, and thus,
no tuning parameters are necessary.
Generally, both approaches involve a large amount of unknown parameters, therefore,  their computation has been quite a challenge.

Efficient algorithms have become indispensable to perform the shape-restricted inference.
\cite{roeneboome.jongbloed:nonpar} provided an extensive discussion on shape-constrained nonparametric inference in their book. In addition, \cite{samworth:recent} gave a thorough review on the shape-restricted log-concave density estimation.
\cite{koenker.mizera:convex}
have proposed a new approach to compute the Kiefer-Wolfowitz nonparametric maximum likelihood estimator in mixtures. In contrast with the prior methods, their new approaches have been cast as convex optimization problems that can be efficiently solved with modern interior-point methods.  \cite{wang.choi:max} proposed some applications of geometric programming in the survival analysis of right-censored covariate data and current status data problems.  \cite{polson.scott.wilard:estrogen} discussed the efficiency of applying proximal algorithms to provide solutions to difficult optimization problems, especially those that involve nonsmooth behavior of the composite objective functions.

The Generalized Additive Models (GAMs)
have been extensively used as a popular dimensional reduction technique in multivariate data analysis over the past two decades.
\cite{hastie.tibshirani:generalized1} and \cite{wood:generalized}, among others, provided comprehensive introductions on spline and GAM-based inference and statistical algorithms.
To combine GAM with shape restrictions,
\cite{chen:generalized} used the active-set algorithm to identify the shape-restricted inference in generalized linear models.
Their method is free of turning parameters and is consistent with the underlying parameters of the compact intervals under mild regular conditions. Moreover, their method is highly competitive with the full parametric regression and reflects an excellent finite sample performance.

In the absence of shape constraints, \cite{hastie.tibshirani:generalized} discussed the generalized additive Cox regression model
applying the local likelihood technique. Given the potential inefficiency of the approach, \cite{chen.ea:global} proposed
a global partial likelihood for nonparametric proportional hazards models. Both methods discussed by \cite{hastie.tibshirani:generalized}
and \cite{chen.ea:global} require the choice of tuning parameter such as window-size, which may not be easy in practice.
In this paper, we study the shape-restricted inference in the additive Cox regression models, and introduce a novel variant of the Cox regression model, namely Shape-Restricted Cox regression (SR-Cox). Recent papers
on shape-constrained Cox regression only cover monotonically constrained covariates \citep{chung.ivanova.el:partial}, whereas the new SR-Cox considers a wide spectrum of shape constraints. There are nine types of generic shapes considered: linear, monotone, convex or concave, and combinations of them. Table 1 comprises a detailed list of the shape constraints.

In SR-Cox, we formulate the underlying estimation of the maximum log-likelihood as a convex optimization problem. \cite{chen:generalized} proposed a similar active-set optimization algorithm to solve the problem, in which the authors successfully applied shape constraints to GAMs.\footnote{The algorithm of the method, the shape-constrained maximum likelihood estimator (SC-MLE) is implemented in the \texttt{R} package \texttt{scar}.} The two main highlights of the algorithm are as follows:
\begin{itemize}

\item The conversion of the SR-Cox regression model into a simple bound-constrained Cox regression through a basis function expansion/transformation. Bound constraints imply that parameters are greater than or equal to zero. This process allows an intuitive management of the exotic types of shape constraints. The basis function expansion is analogous to spline expansion, which approximates a nonlinear curve with a piecewise constant or a piecewise linear function. Given the flexibility of selecting any local knots as expansion points, the approach is categorized as a non-parametric statistical method.

\item The solution of the reformulated bound-constrained optimization problem through an active-set optimization algorithm. The active-set algorithm iteratively solves an optimization subproblem based on a subset of indexes called ``inactive set", which is defined as the index set in which the coefficients are strictly greater than zero; whereas the residual set is called ``active set" and comprises the coefficients that are equal to zero. The neat feature of this method is that working on a Cox regression over the ``inactive" index set at each iteration reduces to a standard unconstrained Cox regression, which can be solved efficiently using an existing Cox regression MLE algorithm. All the Cox regression features such as left truncated, right censoring, or replicated observation time, are inherited.
\end{itemize}

Typically, when applying the basis expansion at the knots, the original problem size becomes considerably larger, which is a common drawback of non-parametric methods.  In Section \ref{sec:active} of the paper, we elaborate on different methods to handle this increasing dimensionality issue.

\cite{chung.ivanova.el:partial} have recently provided a discussion on the Cox regression model with an unspecified monotone covariate
function. Given the Cox partial likelihood as a starting point, they applied
the iterative convex minorant (ICM) sequentially to identify the maximum likelihood estimation of the covariate function. However, this algorithm is not stable given that it attempts to update a large number of parameters simultaneously using a quasi-Newton method
and a quadratic approximation of the log partial likelihood.  Consequently, the algorithm may fail to converge about $10\%$ to $15\%$ of the times. Motivated by \cite{bertsekas:nonlinear} and \cite{hastie.ea:elements}'s popular block coordinate descent algorithm , \cite{qin.et:etrogen} proposed an iterative algorithm that alternately minimizes the model parameters and baseline hazard functions. This algorithm converges to the global maximum. Moreover, in the mixed shape-restricted Cox regression model, in which one covariate is linear and the other is an unspecified monotone function, \cite{qin.et:etrogen} found that the linear regression parameter estimation can be biased when the sample size is small. Therefore, the conventionally used methods such as Jackknife or Bootstrap, are necessary used to correct such bias. In this paper, we have identified that the concave or convex shape-constrained Cox regression is much more stable and efficient than the monotone
shape-constrained Cox regression. No bias correction is necessary even for small sample sizes if convex or concave shape is used.
 Moreover, the results of our simulation study show that even if the shape type is misspecified in the
shape-restricted Cox model, the estimates for the linear regression coefficients are less biased than those derived from the misspecified regression function in the Cox regression model.

The paper is organized as follows. In Section \ref{sec:standardcox}, we briefly describe the setup of the standard Cox regression, with a discussion of all the identification and consistency issues.  In Section \ref{sec:sr-Cox regression} we present the SR-Cox regression model.
In Section \ref{sec:active}, we provide the theoretical framework to solve the SR-Cox regression by applying the active-set algorithm.
 In Sections \ref{sec:example} and \ref{sec:real}, we conduct simulation studies and real data analyses to illustrate the proposed SR-Cox regression, and show that the proposed method generates robust model fitting results and an accurate response function of each covariate.  Finally, in Section \ref{sec:conclusion}, we provide our conclusions, including remarks and several future research directions.

\section{Cox regression introduction}\label{sec:standardcox}

In each data entry, a pair $(\tilde{T},C)$ is defined, in which  $\tilde{T}$ represents the failure time and $C$ is the right censoring time.
The distribution of $\tilde{T}$ can be specified through a mode-specific marginal hazard
\[
\lambda_M(t)=\lim_{\Delta t \rightarrow 0}\frac{P(\tilde{T}<t+\Delta t |\,\tilde{T}\geq t)}{\Delta t}
\]
In the presence of covariates, the conditional version is defined as follows:
\[
\lambda(t|\,x)=\lim_{\Delta t \rightarrow 0}\frac{P(\tilde{T}<t+\Delta t |\,\tilde{T}\geq t,x)}{\Delta t}
\] where $x$ is the input data with dimension $d$.
The Cox proportional model is given by as follows:
\[
\lambda(t|\,x)=\lambda(t)\exp(x\beta),
\]
where $\lambda(t)$ is an unspecified baseline hazard function.

Suppose that $n$ subjects are observed with right possibly censored sample $(t_l,\delta_l, x_{(l)}), l=1,2,...,n$, where
 $t_l=\min(\tilde{T}_l,c_l)$ is
the observed survival time, and $\delta_l=I(t_l\leq c_l)$ is the failure indicator. Note that $\tilde{T}_l$ is censored if $\delta_l=0$. The log-likelihood is
\begin{equation}\label{eqn:likelihood}
\ell=\prod_{l=1}^n \lambda^{\delta_l}(t_l)\exp\left\{-\Lambda(t_l)\exp\left(x_{(l)}^T\beta\right) \right\}
\end{equation}where $\Lambda(t)  = \int_0^t \lambda(u)du $ is the cumulative hazard.

As shown by \cite{breslow:discussion}, with a fixed $\beta$, we only need to consider the $\lambda(t)$ with jumps at each observed failure data point in order to maximize this log-likelihood with respect to $\lambda(t)$. Denote
\[
\lambda_l=\lambda(t_l), l=1,2,...,n_1
\]
as the jump sizes, where $n_1=\sum_{i=1}^n\delta_i$. After discretizing $\lambda(t)$, \cite{breslow:discussion} shows that the maximum  profile
log-likelihood is equivalent to the celebrated Cox's (\citeyear{cox:reg, cox:partial}) partial log-likelihood function

$$\ell_{partial} = \sum_{i=1}^n \delta_i \left\{x_i\beta-\log\left[\sum_{j=1}^n\exp(x_j\beta)I(t_j\geq t_i)\right]\right\}.$$

\subsection{The Cox regression model with additive regressors }

In addition to covariate $X=(X_1,X_2,\ldots,X_{d_x})$, suppose we can also collect covariate $Z = (Z_1, Z_2, \ldots, Z_{d_z})$,
an natural extension of the Cox regression model to the additive Cox regression model is
\[
\lambda(t|z,x)=\lambda(t)\exp\left\{z\beta^z+\sum_{i=1}^{d_x}r_i(x_i)\right\}
 \]
 where $r_i(\cdot), i=1,2,....,d_x$ are an unspecified function of $x_i$.
 Usually, if there is no additional restriction on $r_i$, the global maximum likelihood estimation does not produce consistent estimates.



In this paper we discuss the shape-restricted maximum likelihood estimation in the additive Cox regression, where
$r_i(\cdot)$ is either monotone increasing, decreasing, concave, convex or a combination of them.
Before we elaborate on the details of the technical algorithms, we will discuss the model identifiability issue and consistency of applying the maximum likelihood
estimation.


First we discuss identifiability issue.

If there are two sets of cumulative hazard functions $\Lambda(t)$ and $\Lambda^*(t)$, and covariate functions
$r_i(x_i), r_i^*(x_i)$  and $\beta^z, \beta^{z*}$ such that
\[
\Lambda(t)\exp\left\{z\beta^z+\sum_{i=1}^{d_x}r_i(x_i)\right\}
=\Lambda^*(t)\exp\left\{z\beta^{z*}+\sum_{i=1}^{d_x}r_i^*(x_i)\right\}
\]
for any $t$ and $x_i, i=1,2,...,d_x$, we get the following: 
\[
\log \Lambda(t)-\log \Lambda^*(t)=z(\beta^{z*}-\beta^z)+\sum_{i=1}^{d_x}\left\{r_i^*(x_i)-r_i(x_i)\right\}
\]
The left hand side only depends on $t$ while the right hand side only depends on  $Z$ and $X$.
As this condition is consistent in any $t$, $z_i$ and $x_i$, the formulas must follow
\[
\log \Lambda(t)-\log \Lambda^*(t)=c
\]
and
\[
\sum_{i=1}^{d_x}\{r_i(x_i)-r_i^*(x_i)\}=c, \beta^z=\beta^{z*},
\]
where $c$ is a constant. 
Based on a similar argument, we conclude that
\[
r_i(x_i)-r_i^*(x_i)=c_i
\]
for any $x_1,...,x_{d_x}$, where $c_i$ is constant and independent of $x_i$, $i=1,2,...,{d_x}$. 
Therefore
\[
\frac{\Lambda(t)}{\Lambda^*(t)}=constant
\]
If we assume that the distribution functions of $x_i, i=1,2,...,{d_x}$ are not degenerated, and
 make restrictions such that
\[
r_i(0)=0, i=1,2,...,{d_x}
\]
 then
 \[
 \Lambda(t)=\Lambda^*(t)
 \]
In other words, the underlying model is identifiable.

Next we discuss consistency.

For the consistency proof, we need the commonly used assumptions for deriving large sample properties of the Cox regression model.

{\bf Assumptions.}

Conditionally on covariate $z_i, x_i$s, the survival function $\tilde{T}$ has an absolutely continuous distribution function $F(t|x)$ with a density function $f(t|x)$. The same assumption applies to the censoring variable $C$. Let $G(c|x)=P(C\leq c|x)$ be its distribution function.
Denote $\tau_H=\inf\{t:\,H(t)=1\}$ as the end point of $H=1-\bar{G}\bar{F}$.

A1). $\tau_H=\tau_G<\tau_F$.

A2).
There exists an $\epsilon>0$ such that
\[
\sum_{|\beta^z-\beta^{z0}|\leq \epsilon}E\left[|z|^2\exp(2z\beta^z)\right]<\infty
\]
and the true $r_i^0(x_i)$s are bounded by some positive constant.

A3). The shape-restricted ML is restricted in the space
\[
\{r,\beta||r_i(x_i)|\leq c, i=1,2,...,{d_x}, ||\beta^z-\beta_0^z||\leq c\}
\]
where $0<c<\infty$.

{\bf Proposition.}
Under the assumptions specified above, the shape constrained maximum likelihood estimation is asymptotically consistent.

We will defer proof of this proposition to the Appendix.


\section{Shape-restricted additive Cox regression model}\label{sec:sr-Cox regression}
In this subsection, we describe the procedure to formulate the new SR-Cox regression model as a convex optimization problem that can accommodate shape restrictions in covariates. We have adopted the algorithm proposed by \cite{chen:generalized} in which the authors apply shape-restricted constraints to GAMs.  Since the log partial likelihood is concave (\cite{chung.ivanova.el:partial}), the convergence of our algorithms is guaranteed.

\subsection{Supported types of shapes}
The SR-Cox regression supports nine types of shape constraints, including linear, monotone, convex and concave, and combinations of these types. These shape types are the same as \cite{chen:generalized}'s list. Table 1 below provides all the nine shape constraints. Each covariate $x_i, i =1, \ldots,{d_x}$ may be subject to a separate shape constraint, as shape constraints are component-wise in each covariate. Even if the model is  marginally convex or concave along each covariate it is not necessary to imply whether the joint effect is convex or concave. In other words, the shape information of each interaction of covariates is not considered.

\begin{table}

\caption{Supported shape constraints}
\label{tab:constraints}
\centering
\begin{tabular}{lll}
\hline
           Shape \# &Shape type  & Shape label   \\ \hline
     	1 &  Linear &  l   \\
       2&  Monotone increasing &  in \\
       3& Monotone decreasing & de \\
	4& Convex & cvx \\
	5& Convex increasing & cvxin \\
	6& Convex decreasing & cvxde \\
	7& Concave & ccv \\
	8& Concave increasing & ccvin \\
	9& Concave decreasing & ccvde \\ \hline
\end{tabular}

\end {table}

\subsection{Constructing SR-Cox regression by basis function expansions}
In SR-Cox, we estimate the non-parametric spline function $r_i(x_i)$ subject to the shape restriction type $q_i$ selected from the prior shape type set $q_i \in \{1,2,\ldots,9\}$ (Table  1). For example, by restricting $r_i(x)$ to be a monotonic increasing function with a shape type \#2 - ``Increasing". First, we define function $f$ as follows:
\begin{equation} f(z,x,\beta)= z\beta^z + \sum_{i=1}^{d_x} r_i(x_i, \beta_i^x)\end{equation}
where $\beta = (\beta^z, \beta^x)$ and $z$ has dimension $d_z$. Note that for the shape type \#1 -``Linear", the function $r_i(x_i, \beta^x_i)$ degenerates to the linear term $x_i \beta^x_i$.

The key step of the tackling process in these exotic shape constraints is applying expansion over some basis functions at a pre-determined knot set. The spline function $r_i(x_i, \beta^x_i)$ is then approximated by stacking all the basis functions with weights, which results in an approximation function that is either a piecewise constant or a piecewise polynomial function. We have described the process in detail below. As a result, the problem of conceptually complex nonlinear Cox regression with shape constraints is converted to a standard linear Cox regression with bound constraints (coefficients $\beta^x \geq 0$).

The procedure to construct the SR-Cox regression is described below. The model requires the input of a candidate knot set, from which the optimization algorithm selects the knots that will be used by assigning a strictly positive weight to the basis function. In actual practice, there are several choices for the knots which we will discuss in the following subsections. Let $\{X_{j,i}\}_{j=1}^{K_i}, i = 1,\ldots, d_x$ be the $K_i$ knots of basis expansion for covariate $x_i$. Assuming the nodes are already ordered for each $i$,
$$X_{1,i} \leq X_{2,i} \leq \ldots \leq X_{K_i, i}$$

\begin{itemize}
\item Using ordered statistics $\{X_{j,i}\} = \{X_{(j),i}\}$. We assume all the points are used as knots, thus $K_i = n$.

\item Using predetermined quantiles, for example $\{X_{j,i}\} = \{0, 0.1, 0.2, \ldots, 0.9,1\}$ quantiles of input $x_i$. If the same set of $K$ quantile thresholds are applied for all dimensions, then, $K_i = K$.

\item Customized knots in the domain of $x_i$.
\end{itemize}

We comment that the choice of all points of observed covariates as knots is the most natural and objective one.
The beauty of active set algorithm is that it can eliminate some of those superfluous knots automatically. This is particular important in medical
applications where the available sample sizes range between a few hundreds to thousands.
The choice of using predetermined quantiles as knots is mainly recommended for large sample size problems, where
the available sample sizes range between a few millions to tens or hundreds of millions.
It would be a big burden computationally if one chooses all order statistics as the knots in those situations.
The third choice of customized knots needs prior knowledge, one may use this cautiously.

Depending on the shape type $q_i$, the individual basis function linked to the knot set $\{X_{j,i}\}_{j=1}^{K_i}$ is defined as
\begin{equation}
 g_{ji}(x_i) =
  \begin{cases}
   \bm{1}_{\{X_{j,i} \leq x_i\}}       & \text{ if } q_i = 2 \text{ (``in")},\\
   \bm{1}_{ \{x_i <X_{j,i} \}}       & \text{ if } q_i = 3 \text{ (``de")},\\
   (x_i - X_{j,i})\bm{1}_{\{X_{j,i} \leq x_i\}}     & \text{ if } q_i = 4 \text{ (``cvx") or } q_i = 5\text{ (``cvxin")},\\
   (X_{j,i}-x_i)\bm{1}_{\{x_i \leq X_{j,i}\}}     & \text{ if } q_i =\text{6  (``cvxde")},\\
(X_{j,i}-x_i)\bm{1}_{\{X_{j,i}\leq x_i\}}      & \text{ if } q_i = 7 \text{ (``ccv") or } q_i = 9\text{ (``ccvde")},\\
(x_i - X_{j,i)})\bm{1}_{\{x_i \leq X_{j,i}\}}     & \text{ if } q_i =\text{8  (``ccvin")},\\
  \end{cases}
\end{equation}

\begin{figure}
\centering
\subfigure[][]{%
\label{fig:step}%
\includegraphics[height=1in]{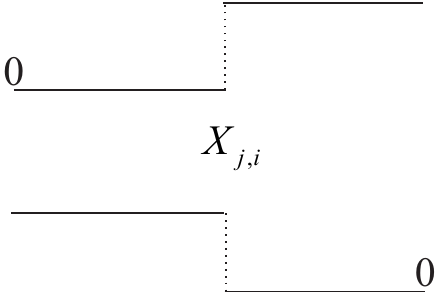}}%
\hspace{100pt}%
\subfigure[][]{%
\label{fig:cont}%
\includegraphics[height=1in]{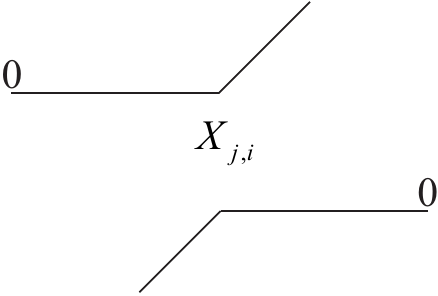}}
\caption{Basis function at knot $X_{ji}$
\subref{fig:step} Step function - monotone increasing (``in") or decreasing (``de") types;
\subref{fig:cont} Continuous function - other convex (``cvx") or concave (``ccv") types;}
\label{fig:basis_fun}%
\end{figure}

Figure~\ref{fig:basis_fun} presents the four basic types of basis functions. Figure~\ref{fig:basis_fun}~(a) plots the monotonic increasing or decreasing types (``in" and ``de"). In this case, the basis function are step functions. Figure~\ref{fig:basis_fun}~(b) plots two (of the four) types of basis functions of the convex or concave type related shapes (``cvx" and ``ccv"). These functions are wedge-shape functions.

When applying basis expansion, $r_i(x_i, \beta^x_i)$ is approximated by a linear combination of the basis functions, as follows:
\begin{equation}
r_i(x_i, \beta^x_i) \sim \hat{r}_i\left(x_i, \{\beta^x_{1i},\ldots,\beta^x_{K_ii}\}\right)  = \sum_{j=1}^{K_i} \beta^x_{ji} g_{ji}(x_i), \text{ subject to: } \beta^x_{ji} \geq 0
\end{equation}

\begin{figure}
\centering
\subfigure[][]{%
\label{fig:cont_app}%
\includegraphics[height=1.4in]{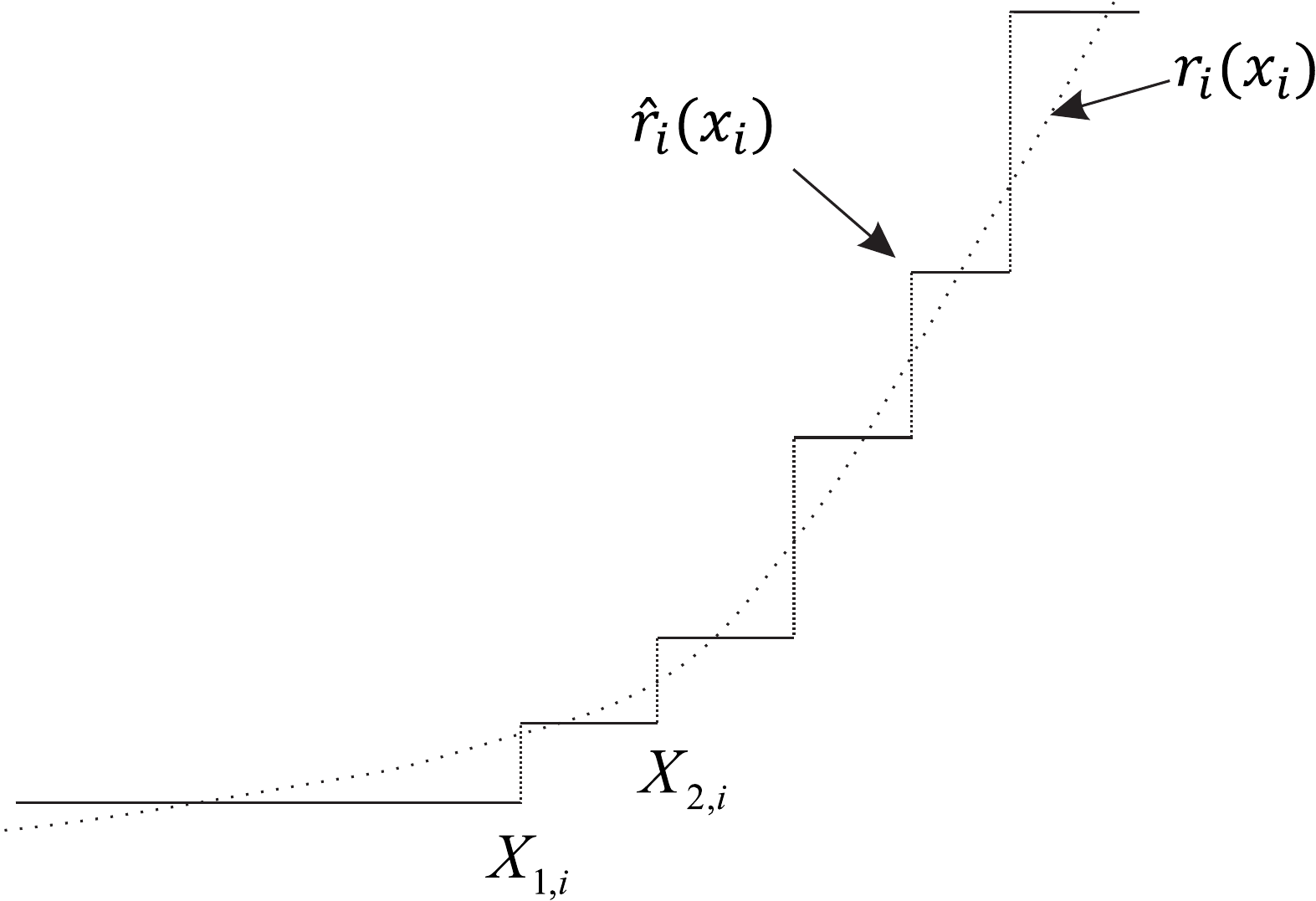}}
\hspace{20pt}%
\subfigure[][]{%
\label{fig:step_app}%
\includegraphics[height=1.3in]{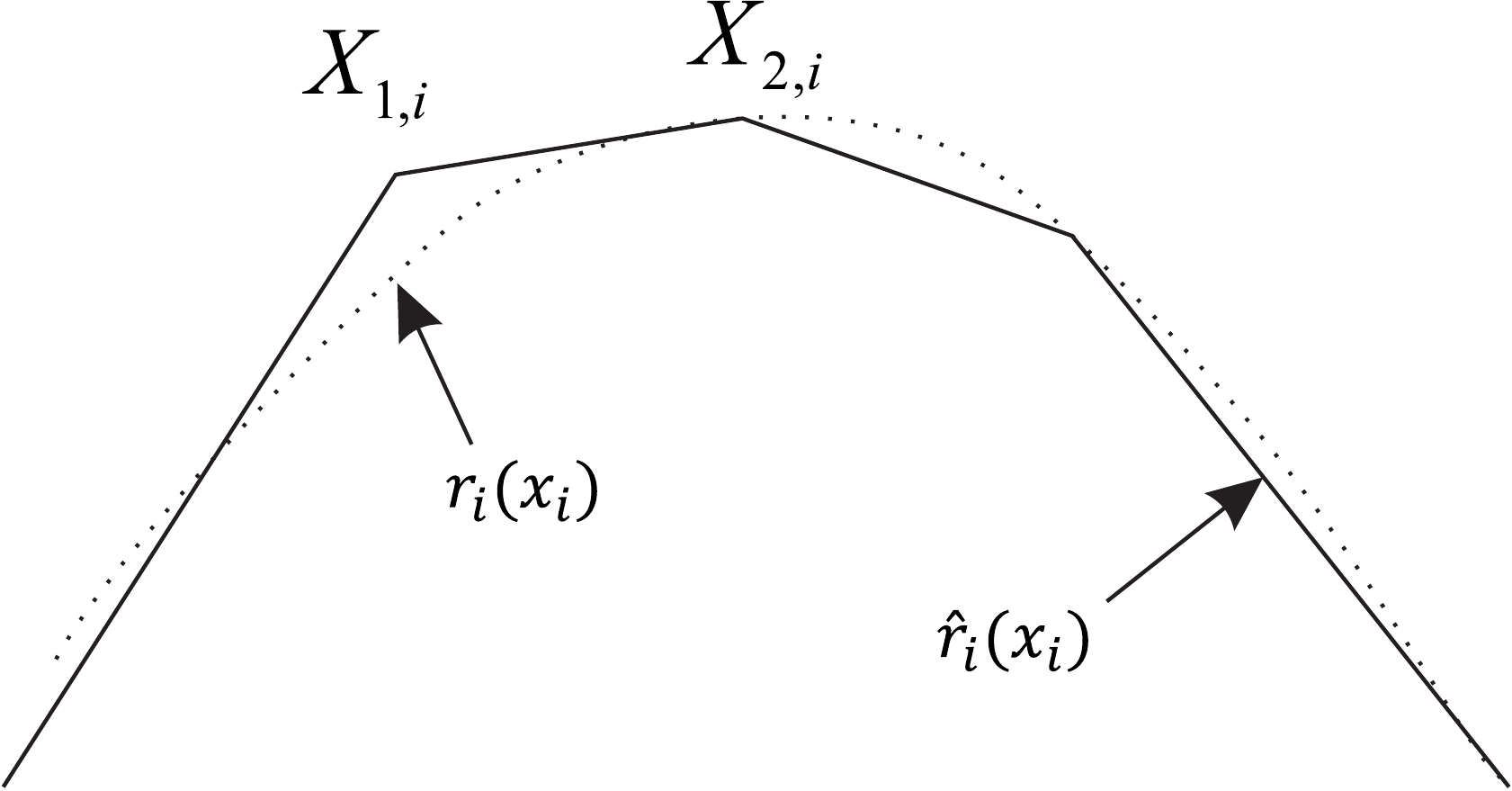}}%
\caption{Approximate nonlinear functions by stacking basis functions
\subref{fig:cont_app} Monotone increasing (``in") case, that is, the sum of basis functions as a piecewise constant function;
\subref{fig:step_app} Concave (``ccv") case, that is, the sum of basis functions as a piecewise linear function;}
\label{fig:basis_fun_app}%
\end{figure}

More specifically, in the monotone increasing or decreasing types, $\hat{r}_i$ is a discrete piecewise constant, as shown in Figure~\ref{fig:basis_fun_app}~(a) below; whereas in the convex and concave types, $\hat{r}_i$ is continuous piecewise linear, as seen in  Figure~\ref{fig:basis_fun_app}~(b). When $\beta^x_{ji}$ is strictly greater than zero ($\beta^x_{ji}> 0$), it indicates that knots $X_{ji}$ are eventually used in the basis function expansion, where there is either a function value ``jump" or a slope change.

The covariates $z_i, i = 1, 2, \ldots, d_z$ are linear type covariates  (type \#1) . The covariates $x_i, i = 1, 2, \ldots, d_x$ are constrained by shape restrictions (type $q_i \geq 2$). If the component function $\hat{r}_i$ is added, the proposed SR-Cox regression optimizes the new function in the form of
\begin{equation}
 \hat{f}(z,x,\beta) =  z \beta^z +\sum_{i=1}^{d_x}\sum_{j=1}^{K_i}\beta^x_{ji} g_{ji}(x_i)
\end{equation}

subject to:
$$
\beta^x \in \mathcal{B} := \begin{cases}
\beta^x_{ji} \geq 0,& \forall j = 1,2,\ldots,K_i \text{ and } q_i\in\{2,3,5,6,8,9\}\\
\beta^x_{ji} \geq 0,& \forall j = 2,\ldots,K_i \text{ and } q_i\in\{4,7\}\\
\end{cases}
$$
As can be seen, all coefficients $\beta^z$ are constraint-free. Compared with the standard linear Cox regression with $d_z + d_x$ parameters, the number of parameters in the problem increases to $d_z+\sum_{i=1}^{d_x} K_i$. More specifically, the parameter vector $\beta = (\beta^z, \beta^x)$ to be optimized is
\begin{equation}
\beta = \{\beta^z_{1},\ldots, \beta^z_{d_z},\beta^x_{11},\ldots,\beta^x_{K_{1}1},\ldots,\beta^x_{1d_x},\ldots,\beta^x_{K_{d_x} d_x}\}
\end{equation}

We formulate the SR-Cox regression as a convex optimization problem, still minimizing the negative partial log-likelihood function. However, the objective function remains the same as the linear Cox regression optimization, as the only difference is that the SR-Cox regression includes additional simple bound constraints $\beta^x \in \mathcal{B}$. Before the conversion of the basis function, the SR-Cox regression is a convex optimization problem, as all the nine shape constraints are convex. However, after the conversion, although we still obtain a convex optimization problem, the bound constraints are quite simpler. For example, in contrast with the original problem in the monotone increasing shape, which requires $r_i(X_{1,i})\leq r_i(X_{2,i}) \leq \ldots \leq r_i(X_{K_i,i})$, the new constraints are based on the selected linear coefficients $\beta^x_{ji} \geq 0$.

The length of parameter $\beta$ increases to $(d_z+\sum_{i=1}^{d_x} K_i)$, which is much larger than that in the unconverted linear Cox regression. Except for the  complex model structure, all the inputs of the Cox regression: events, start/end times, and censoring information are inherited.

\section{Apply active-set optimization algorithm}\label{sec:active}

The SR-Cox regression is formulated as a constrained optimization problem, which can be treated as a standard Cox regression optimization problem with simple non-negative bound constraints. In this section, we present an active-set optimization algorithm to solve the problem.

The active-set algorithm is a widely used constrained optimization algorithm (which also covers simple bound constraints) \citep{nocedal.wright:numerical}. The active-set is defined as the index set that holds the equality constraint conditions. In our optimization setup,  the set $\mathcal{B}$ defines all the bound constraints. The active-set of a feasible solution point $\beta = (\beta^z, \beta^x_{ji})$ includes all the indexes of ``active" constraints   $$\{(j,i)\,|\,\beta^x_{ji} = 0\} $$ The complement set is called an inactive-set and is defined as $$\{(j,i)\,|\,\beta^x_{ji} > 0 \}$$ The active-set algorithms can fall into three categories: primal, dual, and primal-dual. Our algorithm in SR-Cox belongs to the primal category which aims to gradually reduce the objective function at each iteration $k$.

The active-set algorithm consists in minimizing the corresponding objective function as an unconstrained optimization subproblem limited to variables in a working set $\mathcal{S}_k$ at iteration $k$. In SR-Cox, the optimization subproblem is formulated as:
\begin{equation}\label{eqn:coxsub}
\hat{\beta}^{(k)} = \text{arg}\min_\beta \hat{f}(z,x,\beta\,|\, (j,i) \in \mathcal{S}_k) =
z\beta^z+ \sum_{i=1}^{d_x}\sum_{(j,i) \in \mathcal{S}_k}\beta^x_{ji} g_{ji}(x_i) \end{equation}
Or equivalently enforcing $\beta^x_{ji} = 0, \forall (j,i) \notin \mathcal{S}_k$. However, given $\mathcal{S}_k$, the optimization subproblem $\min \hat{f}$ reflects the exact same form as a standard linear Cox regression. This allows us to solve the problem with any standard Cox regression solver efficiently and without getting into the algorithmic complexions in the solvers.

Another key use of the active-set algorithm is to update the working index set $\mathcal{S}_k$ at each iteration $k$. The algorithm process may be performed in the following two ways:
\begin{enumerate}
\item If the new solution $\hat{\beta}^{(k)}$ to the subproblem is feasible,  the current iterate is set as the new solution $\beta^{(k)} =\hat{\beta}^{(k)}$. The algorithm then checks the optimality condition of the main problem. If the optimality condition is not met, algorithm updates the working index set $ \mathcal{S}_k$ by adding a new index. Among all the choices available to select a new index, we select the index corresponding to the maximum gradient function of the objective.

\item
If the new solution is infeasible, the algorithm performs a linear interpolation of the previous iterate  $\beta^{(k-1)}$  to the new solution $\hat{\beta}^{(k)}$  to obtain a feasible iterate  $\beta^{(k)}$ that is exactly constrained by one of the boundary conditions. The intention is to generate a new iterate while the objective function still decreases. The working set is updated by removing the newly triggered active index.
\end{enumerate}

Given that the objective function decreases at each iteration and the number working set combinations is finite, the algorithm converges to the optimal solution in a finite number of iterations.

We formalize the active-set optimization algorithm in the following four steps:

\begin{enumerate}
\item[Step 1:]  The starting working set is initialized as $\mathcal{S}_1 = \{(0,i)\,|\, i=1,2\ldots,d_z\} \cup \{(1,i) |\, q_i = 4, 7\}$

\item[Step 2:] At $k$th iteration, solve the active-set subproblem (\ref{eqn:coxsub}) to obtain a potential solution $\hat{\beta}^{(k)}$.

\item[Step 3:] If the solution $\hat{\beta}^{(k)}$ is infeasible $\hat{\beta}^{x(k)} \notin \mathcal{B}$, implying some components are negative),  a step size multiplier is applied to map the solution back to the feasible domain. Determine a maximum ratio $p\in [0, 1]$ such that the interpolated iterate $\beta^{(k+1)} = (1-p) \beta^{(k)}  + p\hat{\beta}^{(k)}$ is feasible. The ratio $p$ is effectively computed as
$$p = \min_{(j,i) \in \mathcal{S}_k\backslash \mathcal{S}_1} \frac{\hat{\beta}_{ji}}{\hat{\beta}_{ji} - \beta_{ji}},  \quad \mathcal{S}_\_ = \text{argmin}_{(j,i) \in \mathcal{S}_k\backslash \mathcal{S}_1} \frac{\hat{\beta}_{ji}}{\hat{\beta}_{ji} - \beta_{ji}} $$

$\mathcal{S}_\_$ represents the active index such that the new solution $\beta^{(k+1)}$ hits the boundary. Then remove the index from the working set $\mathcal{S}_k:=\mathcal{S}_k\backslash \mathcal{S}_\_$ and proceed to Step 2 to rerun the subproblem.

\item[Step 4:] The iterate  $\beta^{(k)}$ should be feasible at this point. Compute the gradient of the primal function $D^{(k)}_{ji}=\frac{\partial \ell}{\partial \beta_{ji}} (\beta^{(k)})$. If the gradient is zero, the optimal solution is obtained; otherwise, compute the maximum index $\mathcal{S}_+ = \text{argmin}_{(j,i)} D^{(k)}_{ji}$ and add it to the working set $\mathcal{S}_k:=\mathcal{S}_k \bigcup \mathcal{S}_+$.

\end{enumerate}

In Step 4, the derivation of the gradient function of the objective function with respect to the parameter $\beta$ follows the score function
$$\bm{D}=\frac{\partial \ell }{\partial \beta} $$
To calculate this gradient, the score function, which is a byproduct of the Cox regression optimization algorithm should be extracted.

In the process described above, whenever an optimization subproblem is solved (Step 2), or an infeasible solution is mapped back to a feasible region by reducing the step size (Step 3), or when the problem constraint set is relaxed by adding a new index (Step 4), the objective function value decreases iteration by iteration. In Step 4, the aim of selecting the maximum derivative index is to find a direction for the objective function to decrease  as sharply as possible.

The efficiency of the active-set algorithm depends on the size of the subproblem and the number of simple constraints. The number of constraints is highly related to the knots selected. As mentioned, the candidate knot set may be selected through the following three common practices.
\begin{enumerate} \item Using the full-ordered statistics, which means that the basis function may bend at almost any knots, but requires exhaustive local search. This method is highly time-consuming, especially when the sample size $n$ is large (in this case, $K_i = n$).
\item Using the quantiles as knots, which is the recommended method in practice. One may typically start with ten quantiles, so that the basis function is flexible enough to curve but without the requirement of additional computational resources.
\item Using the third approach with customized knots, which requires a high understanding of the relationship between specific covariates and the prediction.
\end{enumerate}

At the beginning of the algorithm, the iteration number $k=1$, that is, the size of the optimization subproblem is small, given that the inactive index set $\mathcal{S}_1$ (the working index set) begins with a small number of indexes, which mainly consists of all the linear constraint indexes. Alternatively, in data input corresponding to the active index set where $\beta_{ji} =0$, the data is screened out of the optimization subproblem. As the algorithm iterates,  new indexes are included in the inactive set, and the computational time increases gradually.

The number of constraints determines the number of iterations of the active-set algorithm. Results show that the total number of iterations used is typically up to 1/2 of the number of parameters in the problem, or $(d_z+\sum_{i=1}^{d_x} K_i)/2$. Additionally, the number of iterations is generally equal to the size of the ``inactive" set, which means that a new index is added to the ``inactive set" in every iteration. Again, this parameter depends on the way of selecting candidate knots (and the number of knots $K_i$ in each covariate $i$). Overall, this is a non-parametric method which is both subject to model flexibility and computational time.

Another relevant topic is the subtle difference between using the ``cvxin" or ``ccvin" shapes or the ``in" shape (and vice versa for the ``de" type). The increasing shape ``in" seems to be a more intuitive setting. However, based on the underlying expansion using basis functions, in Figure~\ref{fig:basis_fun_app}, approximation of the monotone shape types actually uses a piece-wise constant function, whereas in other convex or concave types, the approximating function is a piecewise continuous function. A piecewise continuous function significantly improves the approximation accuracy and even applies fewer knots. In addition, it requires fewer optimization iterations and the algorithm has a faster convergence rate. Another drawback of using the piecewise constant approximation is that it may cause an overfit in the data and more bias in the two tail regions. From our experience, ``cvxin" or ``ccvin" types of shapes are more preferable to handle increasing shape constraints than the standard ``in" shapes. Even though choosing convex vs.~concave may be intriguing at the beginning, based on in real examples, the use of the opposite type of shape (i.e., convex types for concave data) results in a degeneration of algorithm to a linear line without the application of knot transformation. Section \ref{sec:example} includes examples in this regard.

\section{Simulation study}\label{sec:example}
We first ran simulation studies to examine the goodness of model fit of the SR-Cox regression. We denote the parameter estimator as $\hat{\beta}$ and  included the standard Cox regression estimator, which is an un-transformed linear estimator, for comparison purposes.

Suppose the hazard function depends on two covariates  $z$ and $x$  ($d_z =1$ and $d_x=1$). We independently generated $x$ from the exponential distribution $\text{Exp}(1)$ or the normal distribution $N(0,1)$, and $z$ from the normal distribution $N(0,1)$. The survival times were generated from a Weibull distribution with a shape parameter of 2 and a scale function of $\text{exp}(z\beta_{z}+r(x))$. The survival time was further right censored by a threshold time from a uniform distribution U(0, 5).  Therefore, the hazard function has the following form:
\[
\lambda(t|\,z,x)=\lambda_{0}(t)\exp( 2z\beta^z + 2r(x))
\]



Let $\beta^z = -1$, so that  $2z\beta^z=-2z$. We also tested the function forms of $r(x)$ to be either a linear or a nonlinear function. When $r(x)$ is a nonlinear function, the standard Cox regression can introduce a clear bias in the estimator $\hat{\beta}^z$ of the linear component $z$ due to the incorrect estimation of $r(x)$. In such cases, SR-Cox is able to recover the true coefficient more precisely.

Table 2 below lists the simulation settings of seven experiments. In Experiments 1 and 2, we used the same nonlinear form of $r(x)=-3\text{log}(x)$, which is a convex decreasing function. The difference between the two experiments lies in the shape constraint applied in the SR-Cox regression. In Experiment 1, the shape constraint is correctly specified as ``convex decreasing", whereas in Experiment 2, the shape constraint is only specified as ``decreasing".  As for Experiments 3 and 4, we applied the function $r(x)=-x^2$ which is a concave function. Whereas Experiment 3 is set with the correct ``concave" shape constraint, and  Experiment 4 is set with the opposite shape constraint ``convex".  As for Experiment 5 the function applied is $r(x)=-|x|$ and SR-Cox is set with the correct shape constraint.  Finally, Experiments 6 and 7 use the ``linear" form of $r(x)=-2x$ which we analyze the possibility of an SR-Cox model overfitting by setting the suboptimal shapes as ``concave" in Experiment 6 and ``decreasing" in Experiment 7.

In the simulation test, we selected the different sample sizes $n=100$, $500$, and $1,000$ to evaluate the model stability. Table 3 lists the mean and standard deviations calculated in the $1,000$ simulation replications.

Table 3 provides the estimates of $2\hat{\beta}^{z}$ in the seven experiments in both the standard Cox regression and the proposed SR-Cox regression. In the cases in which $r(x)$ is non-linear and the SC-Cox regression correctly specifies the shape constraint (for example, in Experiments 1, 3, and 5), the standard Cox regression generates an estimate $\hat{\beta}^{z}$ with large bias. However, the SC-Cox regression significantly improves the accuracy of the estimation.  In the cases in which $r(x)$ is nonlinear and SC-Cox specifies the shape (for example, in Experiment 2) partially correctly, SC-Cox performs better than the standard Cox, but still shows some estimation bias. As for the cases in which $r(x)$ is non-linear and SR-Cox incorrectly specifies the shape (for example, in Experiment 4), both the standard Cox regression and SC-Cox yield the exact same estimates.  In other words, if the constraint is not specified correctly, $r(x)$ reverts to the linear form, which makes the regression equivalent to the standard Cox regression. In Experiments 6 and 7,  the underlying function is set as a decreasing linear function. Setting the shapes as ``concave" and ``decreasing" also generates accurate parameter estimations comparable to the standard Cox regression. The ``concave" shape setting slightly outperforms the ``decreasing" shape setting which reflects some bias due to the deficiency in the approximation of the  step function approximation in the tail region.

As seen in Table 3, as the sample size increases, the standard deviation of the estimates decreases, which is expected. When the sample size increases from 100 to 500, the estimation accuracy also improves significantly. For example, in Experiment 1, in which the mean estimate of $2\beta^{z}$ is -2.1029, introducing a bias of 0.1029 when $n=100$, but it decreases to 0.041 when $n=500$. When changing the sample size from 500 to 1000, the estimates accuracy only shows a marginal improvement, while the standard deviation decreases (precision increases).

\begin{table}

\caption{Numerical simulation setting for Experiments 1-7. $2z\beta^z=-2z$, $z\sim$ Norm(0,1)}
\label{tab:simu-setting}
\centering
\begin{tabular}{rrrrrr}
  \hline
    &  Exp &  $x$ distribution& $r(x)$ & $r(x) $ shape & SR-Cox constraint\\
  \hline

&$1$ &Exp(1) & $-3\log(x)$ & cvxde & cvxde \\
&$2$ &Exp(1) & $-3\log(x)$ & cvxde & de   \\
&$3$ &Norm(0,1) & $-x^2$ & ccv & ccv \\
&$4$ &Norm(0,1) & $-x^2$ & ccv & cvx \\
&$5$ &Norm(0,1) & $-|x|$ &ccv & ccv \\
&$6$ &Norm(0,1) & $-2x$ &l& ccv\\
&$7$ &Norm(0,1) & $-2x$ &l& de\\
   \hline
\end{tabular}

\end{table}

\begin{table}
\caption{Compare SR-Cox with the standard Cox regression. The shape constraints of SC-Cox are consistent with those of the underlying $r(x)$, and the true value is $2\beta^z=-2$.}
\label{tab:simu}
\centering
\begin{tabular}{rrrrrrrrrr}
  \hline
    &    &\hspace{2em}$n=100$ & & $n=500$ && $n=1000$ & \\
Exp & Method  & Mean & Std.  & Mean & Std.  & Mean & Std. \\
  \hline
1 & SR-Cox&-2.1029&0.2873&-1.9559&0.1016&-1.9322&0.0705 \\
2& SR-Cox &-1.5712&0.2501&-1.4676&0.0928&-1.4550 &0.0648 \\
1\&2 & Cox  & -1.2507&0.1856&-1.2248&0.0764&-1.2233&0.0534 \\
3 & SR-Cox&  -2.1693&0.3017&-2.0325&0.1086&-2.0213&0.0769  \\
4 &  SR-Cox & -1.0472&0.2578&-0.9404&0.1067&-0.9288&0.0772   \\
3\&4 & Cox & -1.0472&0.2578&-0.9404&0.1067&-0.9288&0.0772  \\
5 & SR-Cox & -2.1146&0.2773&-2.0164&0.1033&-2.0116&0.0763    \\
5&  Cox  & -1.4148&0.2481&-1.3186&0.1024&-1.3051&0.0775   \\
6 &  SR-Cox  & -2.0727&0.2312&-1.9970&0.0884&-2.0059&0.0681    \\
7 &  SR-Cox  & -2.1043&0.2631&-1.9232&0.0891&-1.9178&0.0685    \\
6\&7 &  Cox & -2.0504&0.2303&-1.9937&0.0874&-2.0035&0.0674  \\
   \hline
\end{tabular}

\end{table}

%


\begin{figure}[p]
\centering
\subfigure[][]{%
\label{fig:x2n100}%
\includegraphics[height=2in]{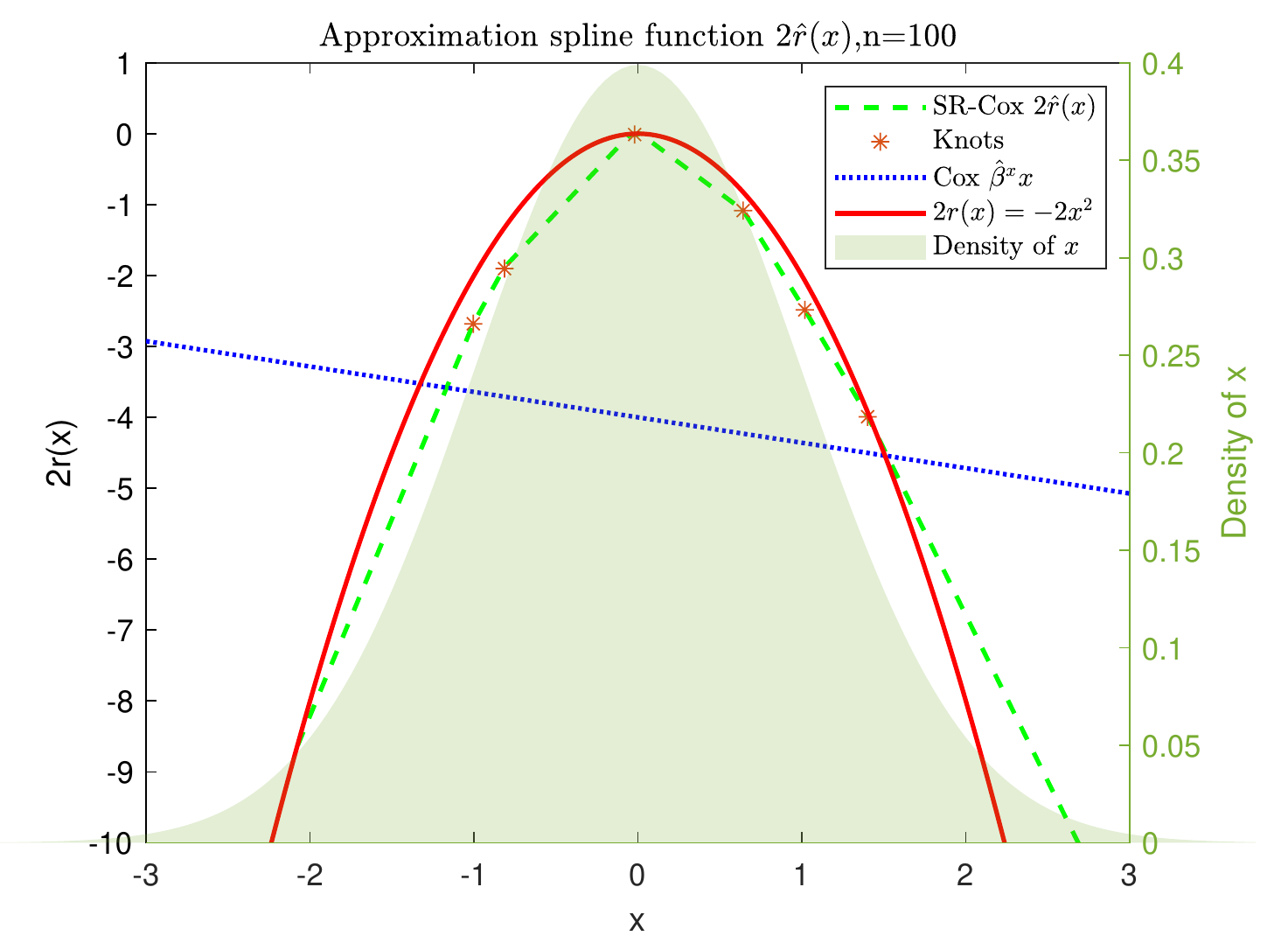}}%
\subfigure[][]{%
\label{fig:x2n500}%
\includegraphics[height=2in]{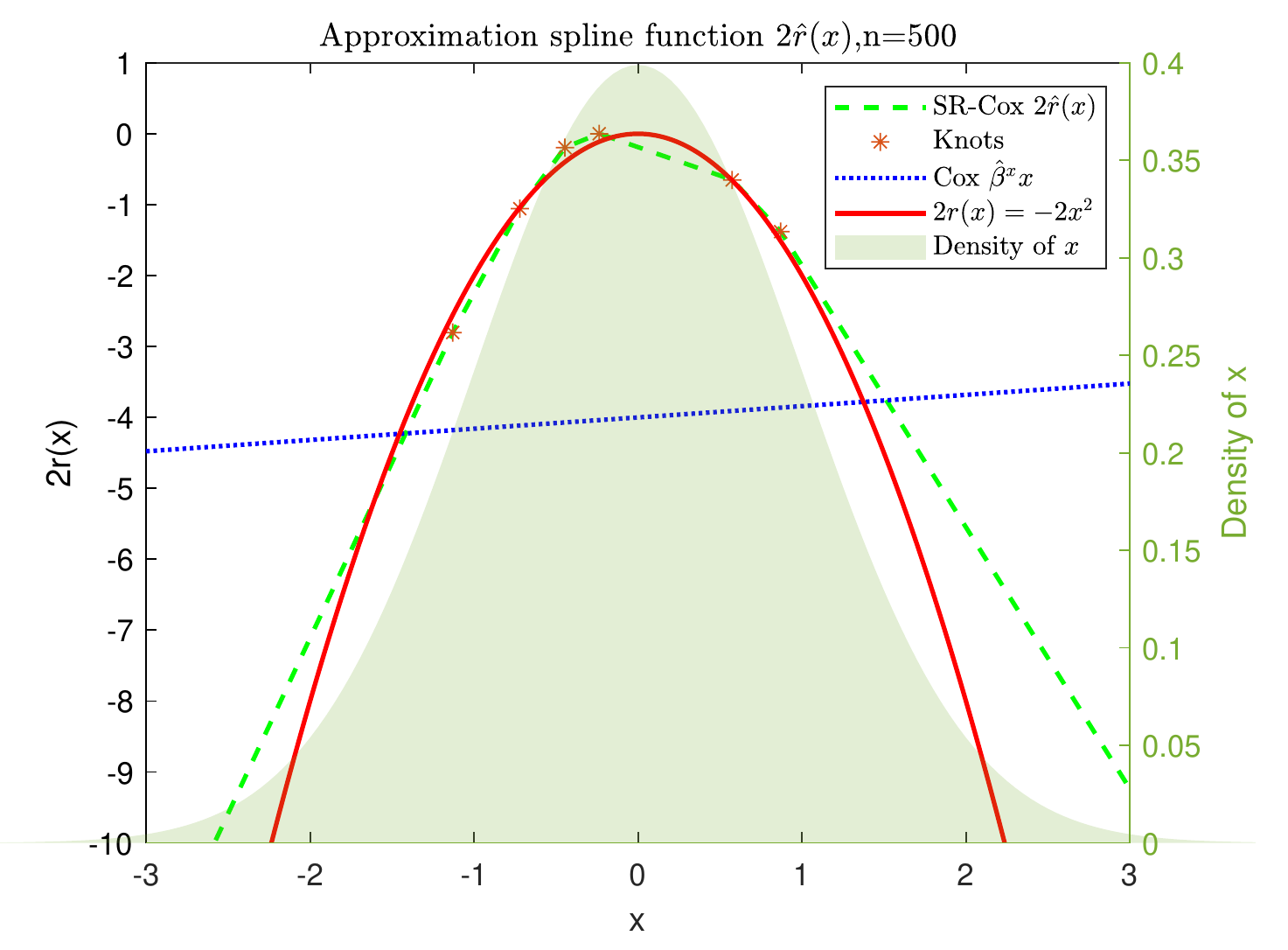}}\quad%
\subfigure[][]{%
\label{fig:x2n1000}%
\includegraphics[height=2in]{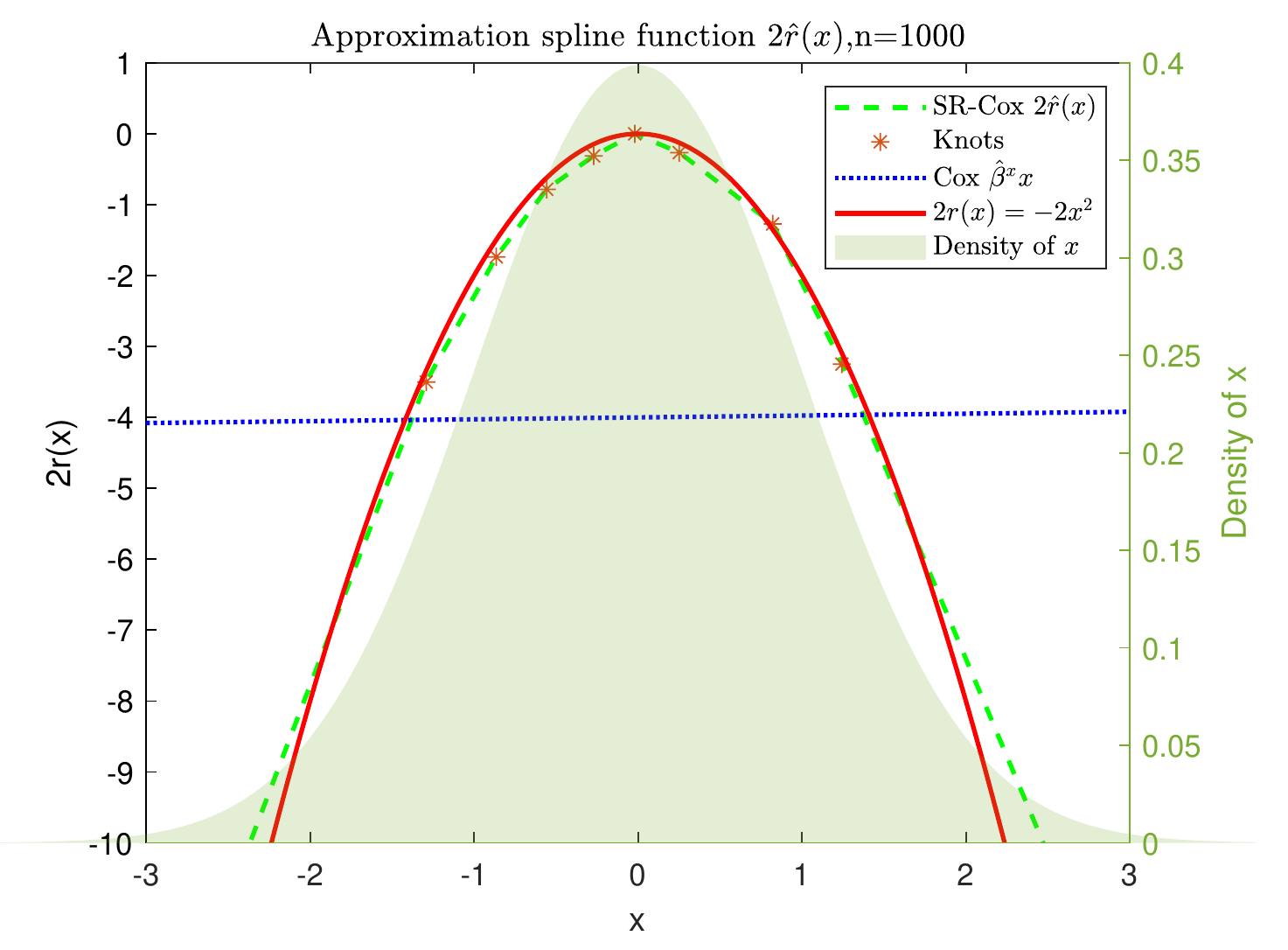}}\quad%
\caption{Piecewise linear spline approximation function $2\hat{r} (x)$ where $2r(x)=-2x^2$ for different sample sizes
\subref{fig:x2n100} n=100;
\subref{fig:x2n500} n=500;
\subref{fig:x2n1000} n=1000;}
\label{fig:fitted_x2}%
\end{figure}

\begin{figure}[p]
\centering
\subfigure[][]{%
\label{fig:logcvxde}%
\includegraphics[height=2in]{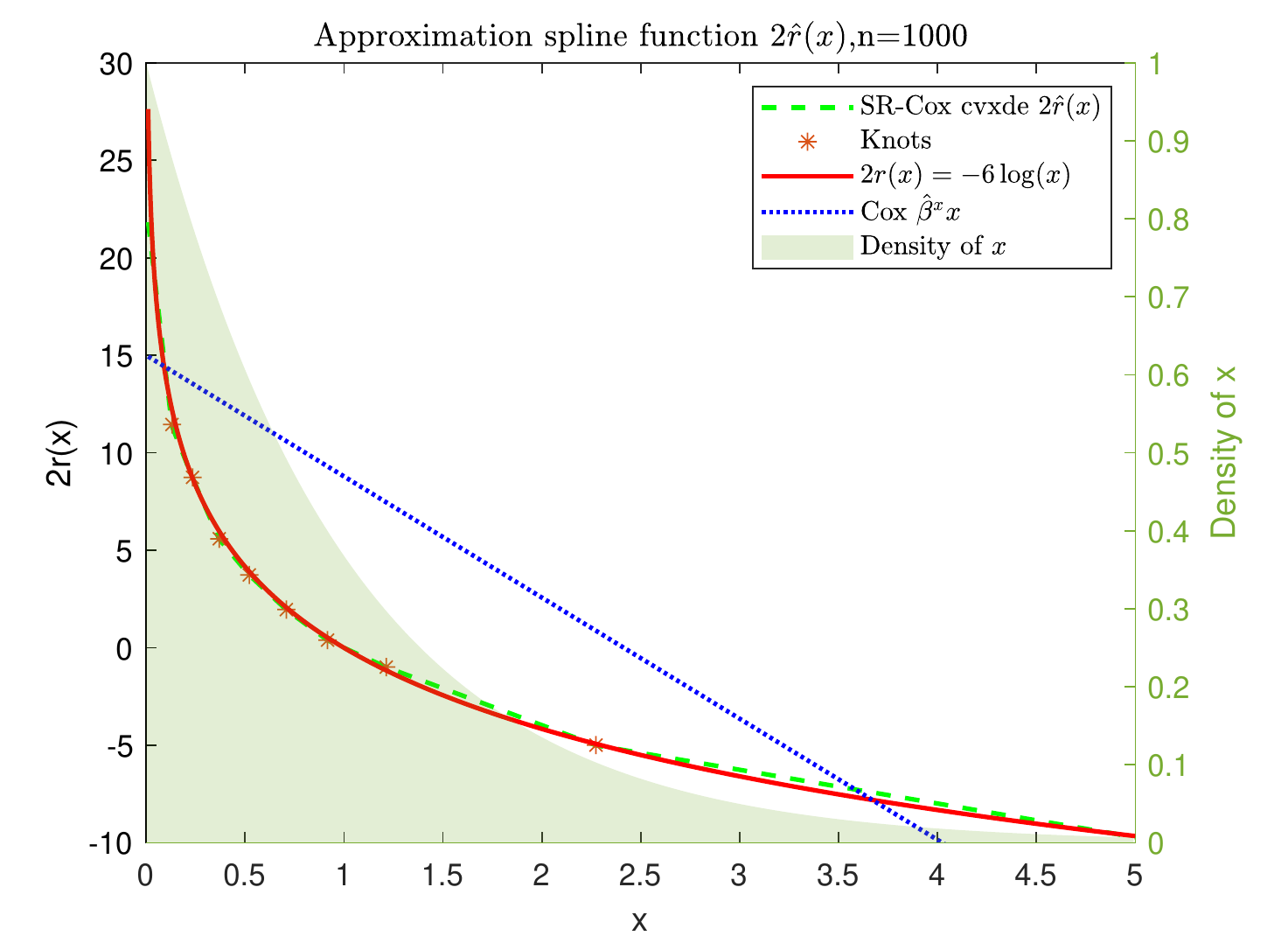}}%
\subfigure[][]{%
\label{fig:logdeq}%
\includegraphics[height=2in]{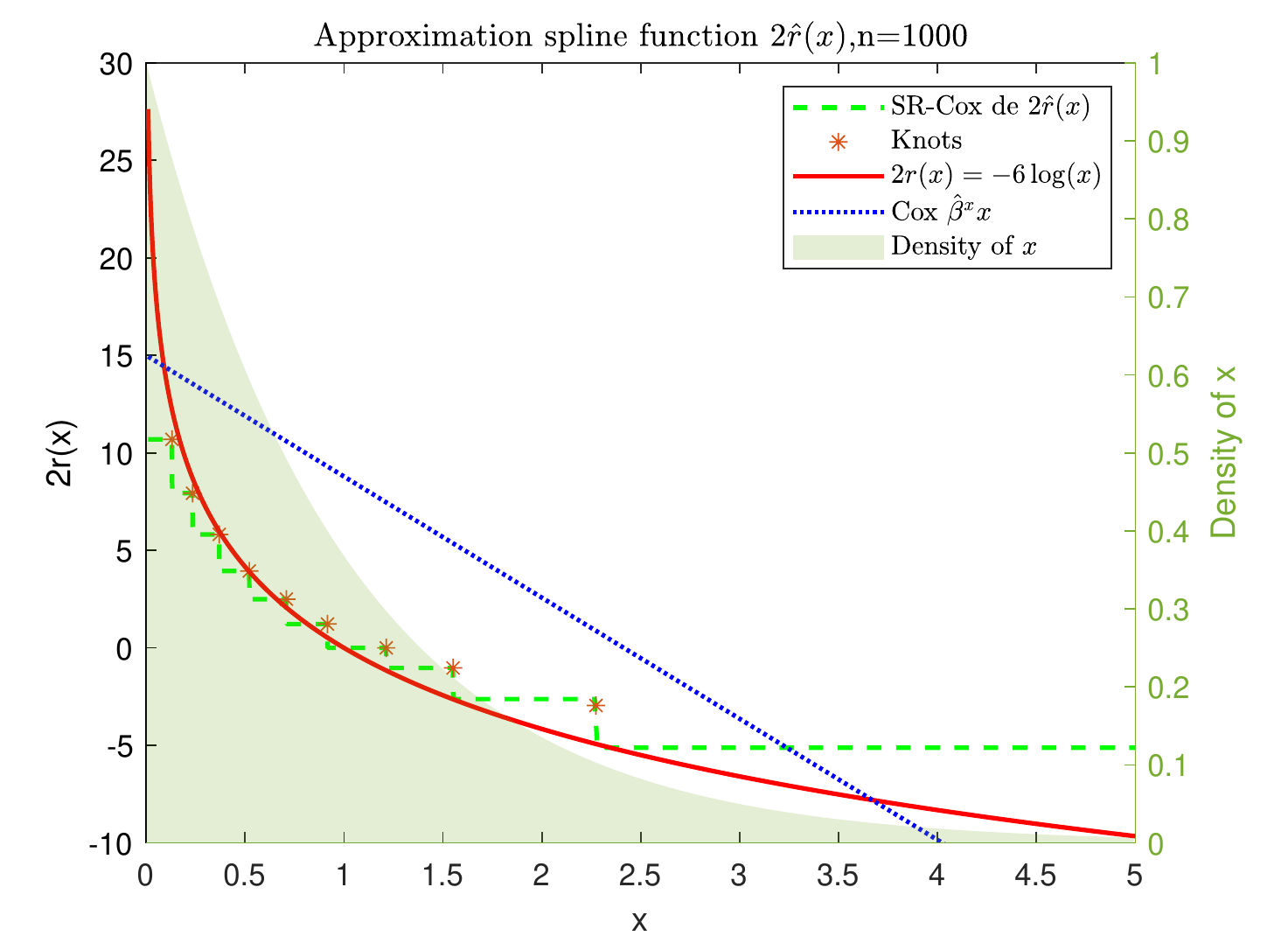}}\quad%
\subfigure[][]{%
\label{fig:logdef}%
\includegraphics[height=2in]{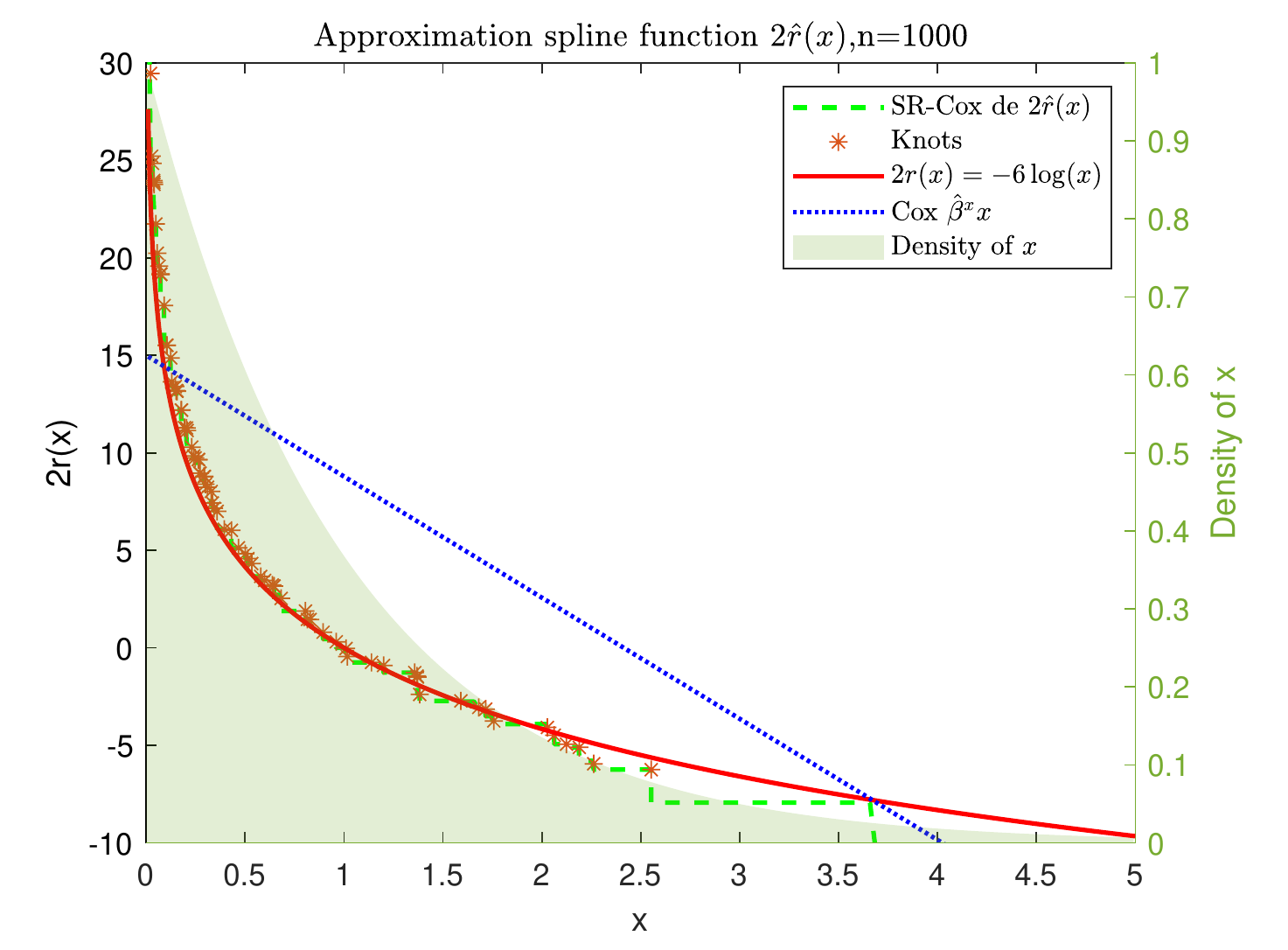}}\quad%
\caption{Piecewise linear spline approximation function vs.\ piecewise constant approximation function $2\hat{r}(x)$ where $2r(x)=-6\log(x)$
\subref{fig:logcvxde} shape=``cvxde" using quantile basis;
\subref{fig:logdeq} shape=``de" using quantile basis;
\subref{fig:logdef} shape=``de" using full ordered statistics basis;}
\label{fig:fitted_logx}%
\end{figure}

Figure \ref{fig:fitted_x2} plots a piecewise linear spline approximation function 2$\hat{r}(x) $ of the underlying $2r(x)=-2x^2$ function in Experiment 3 using one simulation replication. Each subplot represents the different sample sizes $n$ = 100, 500, or 1,000. The red line is the true $2r(x)$; the green dashed line is the fitted spline function from SR-Cox;  the red stars are the knots determined by SR-Cox; the blue dashed line is the fitted 2$\hat{r}(x)$ from the standard Cox regression; and the light green shade shows the normal density function of $x$. Due to the identification issue, a constant shift is applied to align the curves.  The approximation of the  underlying function is quite precise in the interval $(-2,2)$, in which the samples are dense, but it is less accurate at the edges of the $x$ domain.  As the sample size increases, the SR-Cox regression approximates the underlying curve better. However, the standard Cox regression predicts a linear estimation function that completely misses the true shape and is unstable from simulation to simulation.

Figure \ref{fig:fitted_logx} shows the piecewise linear/constant spline function $2\hat{r}(x) $ of the underlying $2r(x)=-6\text{log}(x)$ function in Experiments 1 and 2. Subplot $(a)$  pertains to Experiment 1 in which the shape constraint is ``cvxde". The figure shows the green piecewise linear spline function and the corresponding knots. In contrast, subplots $(b)$ and $(c)$ pertain to Experiment 2 where the shape constraint is ``de". The method applied to select knot set in the spline definition is different in subplots $(b)$ and $(c)$.  In subplot $(b)$, the knots were selected by the quantile method with 10 quantiles, whereas in $(c)$ the full ordered statistics are used as the knot set (1,000 knots). As can be seen, using the full ordered statistics as the knot set results in a very dense selection of knots, which also takes the algorithm much longer (and more iterations) to converge. In addition, at the tail distribution of  the $x$ domain in subplots (b) and (c), the piecewise constant approximation is less accurate than the piecewise linear approximation. This suggests that using the convex/concave types of constraints is more preferable than simply applying the increasing/decreasing shape types. Overall, the estimates of the quantile-based SR-Cox are accurate enough and show computational efficiency.

\section{Real data illustrations}\label{sec:real}

In this section we applied the SR-Cox regression to analyze two real data sets.

\subsection{Mayo clinic primary biliary cirrhosis data}

\cite{fleming:counting} made a data set from the Mayo Clinic trial in primary biliary cirrhosis (PBC) of the liver conducted between 1974 and 1984 available in Appendix D of their monograph.  A total of 424 PBC patients, referred to the Mayo Clinic during that ten-year interval, met the eligibility criteria to participate the randomized placebo controlled trial of the drug D-penicillamine. The first 312 cases of the data set belong to patients who participated in the randomized trial, and thus, their related data is complete. As for the additional 112 cases, the data belongs to patients who did not participate in the clinical trial, but still consented to have basic measurements recorded and be followed for survival. Six of those cases were lost to follow-up shortly after their diagnosis; therefore, the data used in this study is only based on the other 418. The complete data set is available in R database.\footnote{\url{ https://stat.ethz.ch/R-manual/R-devel/library/survival/html/pbc.html}}

By using the Cox's partial likelihood, \cite{fleming:counting} concluded that the best survival model includes explanatory variables such as age, log(albumin), log(bilirubin), edema, and log(protime). Their analysis results are included in Table 4.6.3 of their book.\footnote{In addition, the sixth variable, hepatomegaly, had been independently predictive of survival until the logarithm transformation of bilirubin was introduced. Given that the variable does not include the values of the additional 106 cases, we did not include them in our regression analysis.}
In our analysis, instead of applying the artificial choice of the log transformation, we used the SR-Cox regression imposing the ``convex decreasing" shape restriction on albumin, ``concave increasing" shape on bilirubin, and ``concave increasing" shape on protime, whereas the age and edema variables remained untransformed linear terms (shape = ``linear"). The results of the SR-Cox regression fit are summarized in Table 4 along with the results of both Cox regression models, with and without the log transformations. After removing two observations with missing values in protime, 416 observations were used in the regressions.

\begin{figure}\centering
\includegraphics[height=4in]{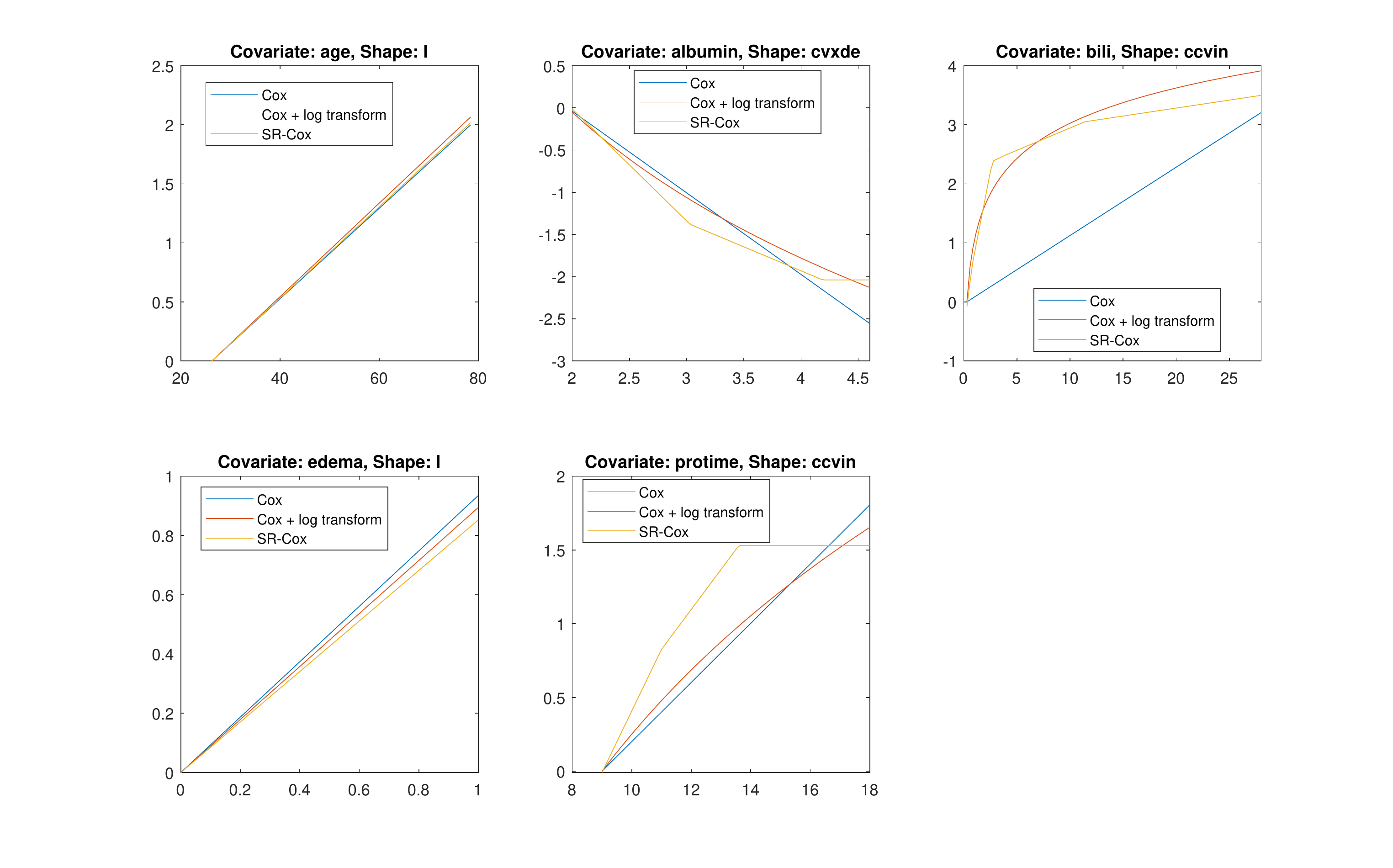}%
\caption{Fitted spline functions of each covariate, including the shape type applied}
\label{fig:biliary}%
\end{figure}

As reported in Table 4, the coefficients of the two linear covariates age and edema are 0.03867 and 0.85255, respectively, which closely match with the Cox regression results. In addition, we applied the likelihood ratio method to obtain the standard deviations of the two estimates. The outputs were comparable to the Cox regressions. Figure \ref{fig:biliary} plots the fitted spline functions of the five covariates (linear and shape-restricted function). As discussed before, the coefficients of the linear terms age and edema are almost the same in the three regressions. The shape-restricted functions are consistent with the parametric log transformation of the variables albumin and bilirubin. In addition, the shape of the spline function of protime is more convex than that of the log transformation, which may be due to the three large protime observations with values greater than 15.

\begin{table}
\caption{Compare SR-Cox model estimates with the standard Cox regressions (with and without log transformations)}\label{tab:pbc}
\centering
\noindent\setlength\tabcolsep{3pt}%
\begin{tabular}{rrrrrrrrr}
  \hline
   Cox &  &  &Cox&  && SR-Cox&  & \\
Var& Coef & Std. & Var & Coef & Std. &Var & Coef & Std.\footnote{Using likelihood ratio method} \\  \hline
age& 0.03832&0.00806& age& 0.03960 & 0.00767 &age&0.03867&0.00816\\
albumin&-0.96822&0.20533 & log(albumin) & -2.49657 & 0.65281&r(albumin)&cvxde&   \\
bili&0.11582 &0.01302& log(bili) & 0.86303 & 0.08295 &r(bili)&ccvin&\\
edema&0.93507 &0.28186 & edema & 0.89460 & 0.27165 &edema&0.85255&0.27806\\
protime&0.20061 &0.05661 & log(protime) &2.38558 & 0.76876 & r(protime)&ccvin&\\
   \hline
\end{tabular}

\end{table}

\subsection{Loan level data study}
In this section, we built a mortgage loan default model using the SR-Cox regression. In the analysis, we modeled the 120-day delinquency rate (defined as the ``default" rate) of the conforming mortgages using loan level data sourced from Freddie Mac's single family loan performance database,\footnote{The public website of Freddie Mac's Single Family Loan-Level dataset is \url{http://www.freddiemac.com/research/datasets/sf_loanlevel_dataset.html}} which we  further merged with macroeconomics variables. Freddie Mac's database comprises loan-level originations, monthly performance, and loss data of fixed-rate single family mortgage loans acquired by Freddie Mac since 1999, which is updated quarterly.

The database is arranged by loan vintage year, that is the origination year. In each vintage year, two data tables - origination and servicing - are stored in the database with the loan ID as the joint field.
\begin{enumerate}
\item	The following loan characteristics are included in the model:
\begin{enumerate}
\item FICO: This is the origination FICO credit score. The higher the FICO score, the higher the chances of borrowers fulfilling their monthly payment obligations and the lower their chances of default. The FICO
score is a measurement of borrowers' credibility.

\item LTV: LTV is a measurement of the level of home equity. High LTV ratios generally indicate a potential risk of borrowers to defaulting on their loan obligations, especially in the event of a drop in the house price. An LTV greater than one implies that the house value is underwater, or that the borrowers' default option is in the money.  We compute the Current LTV (CLTV) based on the  housing price index and the original LTV.

\item DTI: DTI reflects the borrowers' ability to pay the loan. DTI is calculated as the ratio of the total monthly debt payment within the total monthly gross income.  DTI is a measurement of mortgage affordability.   Higher DTIs indicate a higher burden on the borrowers, which translates into a higher probability of default. The origination DTI is available in the origination table.

\end{enumerate}

\item	The servicing table tracks the monthly loan performance with the delinquent status in months. If a loan is performing, the delinquent status code is 0. The 120-day delinquency event ``D120" is flagged when the current month delinquent status = 4 (loan delinquent for 120 days ) and the previous month delinquent status $\leq3$ (loan delinquent for 90 days or fewer).

\item HPI and unemployment rate are the two macroeconomic variables considered in the regression:

\begin{enumerate}
\item HPI: HPI is a broad measure of the movement in the single-family house prices in the United States.  We use the state-level HPI indexes.  A rising housing price market generally creates more job opportunities, which also stimulates consumer confidence and prompts higher spending. If housing prices fall, consumer confidence is eroded  which may potentially trigger an economic recession.  The House Price Appreciation (HPA), derived from HPI,  is negatively associated with the mortgage default rate. In the model, we use the change of HPI, or HPA: $dh=\text{HPI}(t)-\text{HPI}(t-1)$.

\item Unemployment rate: Unemployment rate also has a direct impact on the mortgage default rate.  In the event of a job loss, the mortgagee is at a height of risk of  being unable to  make scheduled monthly payments. Therefore, an increase in the unemployment rate is positively associated with the mortgage default rate. We only use data of the unemployment rate at the state level from the US Bureau of Labor Statistics.\footnote{\url{https://www.bls.gov/}} In the development data, we use the change of unemployment rate: $du=u(t)-u(t-1)$.
\end{enumerate}
\end{enumerate}

In this paper, we applied SR-Cox to the mortgage loans originated in 2007 (2007 vintage year). We selected the year 2007 because it represents the start of an economic downturn period with a relatively large number of default observations. The observation cut off date is March 2019. Though the source servicing data is provided monthly,  the loan data is eventually aggregated at a yearly level; which means each loan is recorded once each year. The attributes of each loan (for example, DTI, CLTV, and FICO) at year $t$ are from the yearly end data of the previous year. The HPI change $dh$ measures the annual HPA, and the unemployment rate change $du$ measures the annual unemployment rate change. We define the D120 event indicator as 1 if there is at least one D120 event in the year. After the data preparation process, there are 4.35 million records in the modeling data of 1.01 million unique loans.  In total, 166 thousand loans ($\approx16.5\%$) ever experienced the D120 event during the life cycle.

\begin{figure}[h]
\centering
\includegraphics[width=6.5in]{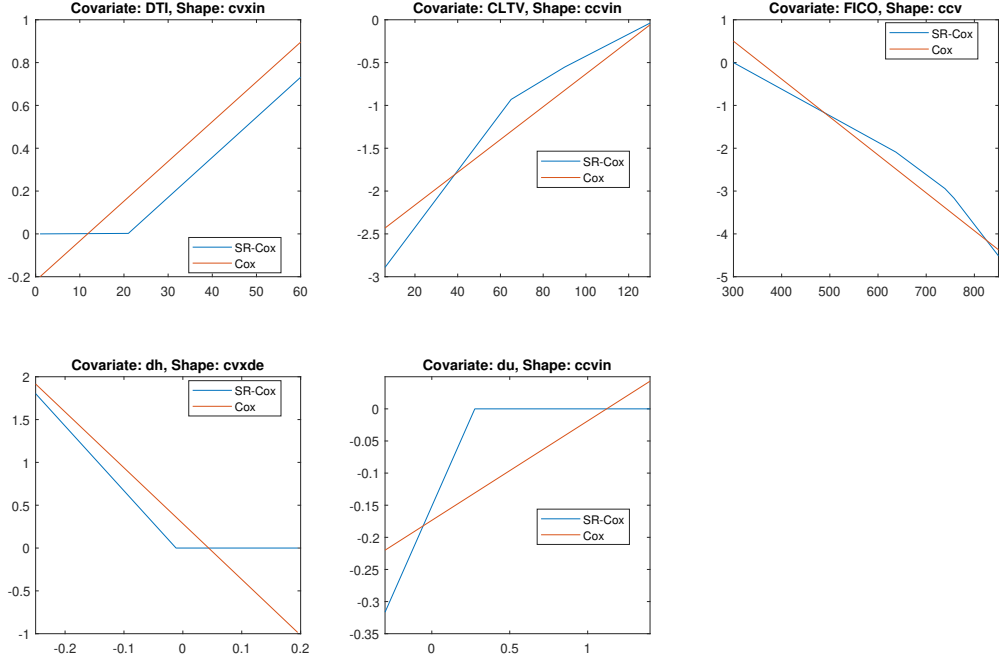}%
\caption{Fitted spline functions of each covariate, including the shape type applied}
\label{fig:fittedsplines}%
\end{figure}

As described, the covariates such as DTI, CLTV, and FICO generally have a monotone relationship with the delinquency rate.  DTI and CLTV are positively correlated with the delinquency rate, whereas FICO has  a negative correlation with it.   We imposed shape constraints on the variables, more specifically: ``convex increasing" on DTI, ``concave increasing" on CLTV, ``concave" on FICO, ``convex decreasing" on $dh$ and  ``concave increasing" on $du$ (Figure \ref{fig:fittedsplines}). In addition, we modeled the loan age by the nonparametric baseline hazard function $\lambda_0(\mbox{age})$ in SR-Cox. The hazard rate funtion in the SR-Cox regression has following form:
\begin{eqnarray}
\lambda(\mbox{age}|\mbox{DTI},\mbox{CLTV},\mbox{FICO}, dh, du)=\\ \nonumber
\lambda_0(\mbox{age})\exp [\,r(\mbox{DTI}, \beta_{dti})+r(\mbox{CLTV}, \beta_{cltv})+r(\mbox{FICO}, \beta_{fico})
+r(dh,\beta_{dh})+r(du,\beta_{du})]\,
\end{eqnarray}
Given the extremely large amount of data in our model, we used the 10\% quantiles as the candidate knot sets, which significantly reduced the computation time. The model training process of the SR-Cox regression took around one hour to complete.

Figure \ref{fig:fittedsplines} below includes the fitted spline functions $r(x)$ of each covariate.  The red line represents the linear effects of standard Cox regression, whereas the blue line represents the fitted shape restricted functions $r(x)$. For DTI, when the ratio is low ($<20\%$), the spline function is a flat line close to zero, which implies that there is virtually zero contribution to default. For loan with DTI higher than $20\%$, the contribution of DTI to default rate increases linearly. The shape of CLTV function is concave increasing, which is similar to the upper part of a logistic function (that is, an ``S-curve"). The curve suggests that at a low range of CLTVs, the default rate increases fast as CLTV increases, which means as CLTV increases to certain level (CLTV = 65), the default rate is less sensitive to the CLTV changes. The FICO component has shown an concave decreasing trending, which suggests that at low FICO scores, the default rate is negatively impacted by the FICO score; while as the FICO score increases higher (especially when FICO $\geq775$), the likelihood of default decreases dramatically. For $dh$ the default rate decreases linearly and then flattens, which implies that decreases of HPI is more relevant to mortgage default, while the impact is small when HPI increases.  Similarly, when the unemployment rate change increases to a certain level, there is a ``burnout-like" effect in the default rate to be flatten. Compared with the standard Cox regression in which all the effects are modeled linearly,  in SR-Cox the non-linear shape approximated by piecewise-linear functions captures the true response to the causes of default more accurately, and offers a more flexible structure to match the business intuitions.

\section{Concluding remarks}\label{sec:conclusion}

The advance in computational algorithms plays an essential role in statistical inference and machine learning research. Various equality and inequality constrained minimization algorithms, such as linear or nonlinear programming, interior point algorithm, active set algorithm, and geometric programming, are applicable to the solution of many  statistical problems. In this paper, we have discussed nine different types of shape-restricted generalized additive Cox regression models. The main attractive feature of our method is that it does not require any turning parameters, which is  crucial, especially in small-sample size problems, since our approach is purely based on data and the selection of knots is objective. In statistical analysis whether or not a log transformation should be applied for the response or a covariate is a thorny issue. 
Clearly a wrong choice of the transformation function in the conventional
Cox regression model analysis may lead to biased results.
Our shape restricted Cox regression model inference can help researchers to determine whether a transformation is necessary or a log transformation is the right choice for the underlying covariate. Methods discussed in this paper are used to analyze a well known clinical trial data set conducted at Mayo clinical center between 1974 to 1984 on primary biliary cirrhosis (PBC) of liver. Our shape restricted inference supports the log transformation for covarites albumin and bilirubin proposed by \cite{fleming:counting} but not for protime.
 Moreover we discuss in details on how to model default rate of a mortgage loan data set with a large sample size of millions.

Overall, the simulation results show that the monotone shape-restricted inference may not bring satisfactory solutions to small-sample size problems. As stated by \cite{qin.et:etrogen}, bias correction methods such as Jackknife or Bootstrap are usually required. However, no bias correction is necessary if the concave or convex shape-restricted is imposed. In these cases, the shape-restricted estimation shows to be almost equivalent to the true parametric estimation (Table 3, Experiments 1, 3, 5, and 6) and achieves excellent results. If the true shape restriction is concave and the working shape restriction is convex,  we end up to the linear shape restriction or the equivalent standard Cox regression model (Table 3, Experiment 4).

Many theoretical challenges such as the local and global convergence of the maximum shape-restricted partial likelihood estimation still require some investigation and will be addressed in future communications.

The algorithms developed in this paper are implemented in Matlab and may be requested at any time.

\section{Appendix}

{\bf Consistency proof}

First we write
\[
\lambda(t|z,x)=\lambda(t)\exp\left\{z\beta^z+\sum_{i=1}^{d_x}r_i(x_i)\right\}=\lambda(t)\exp(z\beta^z)u(x)
\]
where $u(x)=\exp\left\{\sum_{i=1}^{d_x}r_i(x_i)\right\}$. The log-likelihood can be written as
\[
\ell=\sum_{i=1}^n\left\{\delta_i \log \{\exp(z_i\beta^z)+u(x_i)\}+\delta_i\log \Lambda(t_i)-\exp(z_i\beta^z)u(x_i)\Lambda(t_i)\right\}.
\]
Using the same notation as in \cite{van.wellner:weak}, we denote $P_n$ as the empirical measure based on
$(T_i,\delta_i,Z_i,x_i), i=1,2,....,n$.
Let $\hat{\Lambda}(t), \hat{r}, \hat{\beta}^z$ be the shape constrained MLE, i.e.,
\[
P_n\ell(\hat{\Lambda}(t)(\hat{\beta}^z,\hat{r}), \hat{\beta}^z,\hat{r})\geq P_n\ell(\hat{\Lambda}(\beta^z,r),\beta^z,r)
\]
for any shape-restricted function $r$ satisfying $r(0)=0$.

Let $dN(t)=\delta dI(T\leq t)$, and $Y(t)=I(T\geq t)$.  The Breslow baseline estimator is
\[
\hat{\Lambda}(t,\hat{\beta}^z,\hat{r})
=\int\frac{P_ndN(t)}{P_nY(t)\exp(z\hat{\beta}^z+\sum_{i=1}^{d_x} \hat{r}_i(x_i))}.
\]

Define
\[
\hat{\Lambda}(t,\beta^{z0},r^0)
=\int\frac{P_ndN(t)}{P_nY(t)\exp(z\beta^{z0}+\sum_{i=1}^{d_x} r_i^0(x_i))}
\]
Since $\hat{\beta}^z$ is bounded sequence of Euclidean parameters, it has a convergence subsequence such that
$\hat{\beta}^z\rightarrow \beta^{z*}$. Also $\hat{r}_i(x_i), i=1,2,...,{d_x}$ are bounded monotonic functions (or concave functions), by Helly's selection theorem we have a convergence subsequence. For notation convenience, we still use themselves, $\hat{r}_n(\cdot)=(\hat{r}_1(x_1),...,\hat{r}_{d_x}(x_{d_x}))\rightarrow r^*(\cdot)$. Denote
\[
\hat{\Lambda}(t,\beta^{z*},r^*)
=\int\frac{P_ndN(t)}{P_nY(t)\exp(z\beta^{z*}+\sum_{i=1}^{d_x} r_i^*(x_i))}
\]
and
\[
\Lambda^*(t,\beta^{z*},r^*)
=\int\frac{PdN(t)}{PY(t)\exp(z\beta^{z*}+\sum_{i=1}^{d_x}r_i^*(x_i))}.
\]
Clearly $\Lambda^*(t,\beta^{z*},r^*)$ becomes the true baseline cumulative hazard $\Lambda_0(t)$ if $r^*=r^0$ and $\beta^{z*}=\beta^{z0}$, that is
\[
\Lambda^*(t,\beta^{z0},r^0)=\Lambda_0(t).
\]


Let $M\in (0, \tau_H)$.
Define
\[
{\cal H}=\left\{ h\;|\; h=Y(u)\exp\left(z\beta^z+\sum_{i=1}^{d_x} r_i(x_i) \right)\right\},
\]
where $||\beta||\leq c$, $0<u\leq M$ and $r_i(x_i), i=1,2,...,{d_x}$ are monotonic (or concave functions). It is well known its entropy with bracketing for the class of monotonic functions
satisfies
\[
\log N_{[]}(\epsilon,{\cal H},L_2(P))\leq 1/\epsilon
\]
see, for example, Theorem 2.7.5 in \cite{van.wellner:weak} and Lemma 9.35 in \cite{kosorok:intro}, and for the class of convex (or concave) functions satisfies
\[
\log N_{[]}(\epsilon,{\cal H},L_2(P))\leq \epsilon^{-1/2}
\]
for example \cite{gao:metric}.

Moreover for $u\in (0, M]$,
\[
E\left[Y(u)\exp\left(z\beta^z+\sum_{i=1}^{d_x} r_i(x_i)\right)\right]=\bar{F}(u|z,x)\bar{G}(u|x,z)\geq \bar{F}(M|x,z)\bar{G}(M|x,z)>0.
\]
We can show ${\cal H}$ is a Vapnik-Chervonenkis (VC) class. For convenience we denote
\[
r(x)= \sum_{i=1}^{d_x} r_i(x_i).
\]
Moreover,
\begin{eqnarray*}
&&\int \frac{P_ndN(t)}{P_nY(t)\exp(z\beta^z)\exp(r(x))}\\
&=&
\int \frac{P_ndN(t)}{PY(t)\exp(z\beta^z)\exp(r(x))}\\
&+&
\int P_ndN(t)\left \{\frac{1}{P_nY(t)\exp(z\beta^z)\exp(r(x))}-\frac{1}{PY(t)\exp(z\beta^z)\exp(r(x))}\right \}\\
&:=& A_n+B_n
\end{eqnarray*}
\[
|B_n|\leq  \frac{\sup_{0\leq t\leq M}|P_nY(t)\exp(z\beta^z)\exp(r(x))-PY(t)\exp(z\beta^z)\exp(r(x))|}
{P_nY(t)\exp(z\beta^z)\exp(r(x))PY(t)\exp(z\beta^z)\exp(r(x))}.
\]
Therefore $\hat{\Lambda}(t,\beta^{z*},r^*)\rightarrow \Lambda^*(t,\beta^{z*},r^*)$ uniformly in $(0,M)$, where $M\in (0,\tau_H)$.

Next we use the same argument as \cite{murphy.van:max}.

The log-likelihood can be written as
\[
\ell=P_n[\delta\log d\Lambda(t)+z\beta^z+r(x)]-P_n[\Lambda(t)\exp(z\beta^z+r(x))].
\]
Let
\[
d\hat{\Lambda}(t,\beta,r)=\frac{P_ndN(t)}{P_nY(t)\exp(z\beta^z+r(x))},
\]
\[
d\hat{\Lambda}(t,\beta^{z0},r^0)=\frac{P_ndN(t)}{P_nY(t)\exp(z\beta^{z0}+r^0(x))}.
\]
Then
\begin{eqnarray*}
\frac{d\hat{\Lambda}(t,\hat{\beta}^z,\hat{r})}{d\hat{\Lambda}(t,\beta^{z0},r^0)}
&=& \frac{P_nY(t)\exp(z\beta^{z0}+r^0(x))}{P_nY(t)\exp(z\hat{\beta}^z+\hat{r}(x))}\\
&\rightarrow & \frac{PY(t)\exp(z\beta^{z0}+r^0(x))}{PY(t)\exp(z\beta^{z*}+r^*(x))}\\
&=& \frac{P\exp(z\beta^{z0}+r^0(x))}{P\exp(z\beta^{z*}+r^*(x))}=\frac{d\Lambda^*(t,\beta^{z*},r^*)}{d\Lambda^0(t)},
\end{eqnarray*}
uniformly by the law of large sample theory since it is indexed by the class of monotonic functions.

Note that
\begin{eqnarray*}
&&P_n\ell(\hat{\Lambda}(\hat{\beta}^z,\hat{r}),\hat{\beta}^z,\hat{r})-P_n\ell(\hat{\Lambda}(\beta^{z0},r^0),\beta^{z0},r^0))\\
&=& P_n[\delta\log\{d\hat{\Lambda}(t,\hat{\beta}^z,\hat{r})/d\hat{\Lambda}(t,\beta^{z0},r^0)\}]+P_n[z\hat{\beta}^z-\hat{\lambda}(t,\hat{\beta}^z,\hat{r})
\exp(z\hat{\beta}^z+\hat{r}(x)]\\&&-P_n[z\beta^{z0}-\hat{\Lambda}(t,\beta^{z0},r^0)]\\
&=& P_n[\delta\log\{P_nY(t)\exp(z\beta^{z0}+r^0(x))/P_nY(t)\exp(z\hat{\beta}^z+\hat{r}(x))\}]\\
&&+P_n[z\hat{\beta}^z-\hat{\Lambda}(t,\hat{\beta}^z,\hat{r})
\exp(z\hat{\beta}^z+\hat{r}(x)]-P_n[z\beta^{z0}-\hat{\Lambda}(t,\beta^{z0},r^0)\exp(z\beta^{z0}+r^0(x))]\\
&\rightarrow & P[\delta \log \{d\Lambda^*(t,\beta^{z*},r^*)/d\Lambda_0(t)\}+
P[z\beta^{z*}-\Lambda^*(t,\beta^{z*},r^*)
\exp(z\beta^{z*}+r^*(x))]\\
&&- P[z\beta^{z0}-\Lambda^*(t,\beta^{z0},r^0)\exp(z\beta^{z0}+r^0(x)]\\
&=& P\ell(\Lambda^*(\beta^{z*},r^*),\beta^{z*},r^*)-P\ell(\Lambda^0(t),\beta^{z0},r^0).
\end{eqnarray*}
Finally we have
\[
0\leq P[\ell(\Lambda^*,r^*,\beta^{z*})-\ell(\Lambda_0,\beta^{z0},r^0].
\]
It is well known that the Hellinger distance $h^2$ is always smaller than the Kullback-Libeler divergence, therefore, we can conclude
\[
h^2(f(\lambda^*,\beta^{z*},r^*),f(\Lambda_0,\beta^{z0},r^0))=0,
\]
where
\[
f(\lambda,\beta,r)=\lambda(t)\exp(z\beta^z)\exp(r(x))\exp\{-\Lambda(t)\exp(z\beta^z)\exp(r(x))\}
\]
is the density function corresponding to $\lambda, \beta^z$ and $r(x)$.
Finally by the identifiability assumption, we have
\[
\Lambda^*(\beta^{z0},r^0)=\Lambda_0, r^*=r^0, \beta^{z*}=\beta^{z0}.
\]

Thus we have shown that any convergence sequence has a limiting to the true underlying parameters.  This concludes the consistency proof.

\section*{Disclaimer}

The opinions in this paper are strictly those of the authors and do not represent the views of Wells Fargo \& Company, or any of their subsidiaries or affiliates.

\bibliography{Shape_Cox_F} {}
\bibliographystyle{agsm}

\end{document}